\documentclass[reprint,superscriptaddress,amsmath,amssymb,aps,pre]{revtex4-2}

\usepackage{graphicx}
\usepackage{dcolumn}
\usepackage{bm}

\usepackage{todonotes}

\usepackage[utf8]{inputenc}

\usepackage{newfloat,algcompatible}
\usepackage[size=small]{caption}
\usepackage{etoolbox}

\AtEndEnvironment{algorithm}{\noindent\hrulefill\par\nobreak\vskip-8pt}

\DeclareFloatingEnvironment[
    fileext=loa,
    listname=List of Algorithms,
    name=ALGORITHM,
    placement=tbhp,
]{algorithm}
\DeclareCaptionFormat{algorithms}{\vskip-15pt\hrulefill\par#1#2#3\vskip-6pt\hrulefill}
\algblock[Input]{Input}{EndInput}
\algblockdefx[Input]{Input}{EndInput}%
    [1]{\textbf{Input} #1}%
    {}

\usepackage{xcolor}
\definecolor{monbleu}{RGB}{76,114,176}

\usepackage[colorlinks=true,allcolors=monbleu]{hyperref}

\newcommand{\TRUE}{\textbf{true}}
\newcommand{\FALSE}{\textbf{false}}
\newcommand{\RETURN}{\textbf{return}}

\begin{document}

\preprint{APS/123-QED}

\title{On the Uniform Sampling of the Configuration Model with Centrality Constraints}

\author{François Thibault}
\affiliation{D\'epartement de physique, de g\'enie physique et d'optique,
Universit\'e Laval, Qu\'ebec (Qu\'ebec), Canada G1V 0A6}
\affiliation{Centre interdisciplinaire en mod\'elisation math\'ematique, Universit\'e Laval, Qu\'ebec (Qu\'ebec), Canada G1V 0A6}
\author{Laurent Hébert-Dufresne}%
\affiliation{Vermont Complex Systems Center, University of Vermont, Burlington, VT 05405 USA}
\affiliation{Department of Computer Science, University of Vermont, Burlington, VT 05405 USA}
\author{Antoine Allard}%
\affiliation{D\'epartement de physique, de g\'enie physique et d'optique,
Universit\'e Laval, Qu\'ebec (Qu\'ebec), Canada G1V 0A6}
\affiliation{Centre interdisciplinaire en mod\'elisation math\'ematique, Universit\'e Laval, Qu\'ebec (Qu\'ebec), Canada G1V 0A6
}%

\date{\today}

\begin{abstract}
    The Onion Decomposition has recently been shown to provide principled models of complex graphs that better reproduce the sparse networks found in nature, but at the cost of complicated connection rules. We propose a $k$-edge swapping MCMC algorithm to efficiently obtain a uniform sample from the ensemble of simple graphs with a fixed Onion Decomposition and degree sequence. We prove the non-connectivity of the 2-edge swap algorithm for some small graphs, but then provide numerical experiments to show that this non-connectivity is not a problem for 2-edge swap in many practical cases, and likely irrelevant when using $k$-edge swaps with $k>2$. We finish by comparing our null model to other well-known models in the literature, and show that keeping constraints on the meso-scale structures of the Onion Decomposition greatly increases both the structural and functional realism of random graph null model.
\end{abstract}

\maketitle


\section{\label{sec:level1} Introduction}

Random graph models play a critical role in the study of complex networked systems. Process-based models, such as the Watts-Strogatz or Barab\'asi-Albert models, serve to understand the building mechanism of networks, while constraint-based models usually serve as comparison tools between empirical network datasets and a null hypothesis over some specific graph properties. In constraints-based model, graphs need to respect a given topological property, for example the degree sequence, while being maximally random in all other aspects.

While mathematical descriptions of these models can offer great analytical insights, sampling algorithms continue to play an important role, especially when applying these models to empirical network datasets. However, building an algorithm that generates a representative sample of a model and its constraints remains a challenge, even when the constraints are as simple as keeping the degree sequence of the graph~\cite{fosdick_configuring_2018, carstens_switching_2017}. This specific problem refers to the sampling of the \textit{Configuration Model} (CM), which is defined as the ensemble of all graphs with a specific degree sequence.

Historically, sampling the Configuration Model has been approached through two different methods: By using a stub-matching algorithm~\cite{kim_degree-based_2009, genio_efficient_2010, diaconis_sequential_2011} in which the challenge is to obtain an algorithm that results in a uniform sample of graphs without creating self-loops or multi-edges, or by using an edge-swapping Markov Chain Monte Carlo (MCMC) algorithm (see Ref.~\cite{fosdick_configuring_2018} for an extended review of how the edge-swapping method has been developed independently in many fields). In the case of the MCMC algorithms, the central challenge is that the mixing time is generally unknown, although it has been conjectured to be fast (KTV conjecture)~\cite{Kannan_1999_simple}. While mixing remains an open problem, theoretical~\cite{cooper_sampling_2007, cooper2012corrigendumsamplingregulargraphs, erdos_new_2018, GREENHILL20181} and numerical~\cite{milo_uniform_2004, dutta_sampling_2023} results hint that the MCMC method is indeed efficient and can be used to efficiently sample graphs from empirical network datasets.

The Configuration Model is behind many important insights in network science~\cite{MASLOV2004, Stouffer_evidence_2007, Chatterjee_2011, Miller_epidemic_2014, stonge2017susceptible}, but it is limited in its ability to reproduce several critical features observed in empirical network datasets~\cite{orsini_quantifying_2015}. While this limitation has not hindered its use to this day~\cite{Zarei_bursts_2024, Okeukwu-Ogbonnaya_towards_2024, Murphy2024, Clariana2024Quantifying}, repurposing the edge-swapping algorithm to include additional structural properties remains a worthwhile endeavor nonetheless. For example, degree-degree correlations have been enforced through a constraint on a joint degree matrix, both in the case of keeping the constraint as an average over the sample~\cite{newman_2002_assortative, newman_2003_mixing} or as an exact constraint~\cite{Stanton_2012_constructing, czabarka2015realizations, amanatidis_connected_2018}. In the latter, it has been shown that the MCMC is fast mixing~\cite{erdos_2015_decomposition}. We refer to this model as the Correlated Configuration Model (CCM). Similarly, there has been work on keeping a constraint on the connectivity~\cite{stauffer_study_2011, ring2020connected} of the network as well as on the clustering coefficient~\cite{jamakovic_how_2015, orsini_quantifying_2015}.

In recent years, many new MCMC algorithms have been proposed to preserve even more complex structures in the graphs: Cooper et al. have worked on an edge-swapping algorithm that produces more triangles than expected at random~\cite{cooper_triangle_2023}; Stamm et al. use constrained swapping to preserve neighborhood structure based on the Color Refinement algorithm~\cite{Stamm_2023_neighborhood}; Van Koevering et al. use an approach distinct from edge swapping to sample graphs with a fixed $k$-core sequence instead of a fixed degree sequence~\cite{van_koevering_random_2021}. 

Our work is inspired by this new focus towards constraints on a mesoscopic scale, and is closely related to the two latter models. We propose a Configuration Model with an added constraint on the Onion Decomposition~\cite{hebert-dufresne_multi-scale_2016}, a refinement of the well-known $k$-core decomposition~\cite{kong_kmathmi_2019}. Our model can be seen as an addition of the model in~\cite{van_koevering_random_2021}, since our ensemble is a more constrained subset of theirs. Additionally, since the calculation of the Onion Decomposition is related to neighborhood structures, our model can be expected to have similarities with the approach of Ref.~\cite{Stamm_2023_neighborhood}.

We first give a definition of the Onion Decomposition and of our Layered Configuration Model, then propose an edge-swapping algorithm to sample from it. We show that it is disconnected when using swaps with only two edges, but provide a means of shuffling the graphs with $k$-edge swaps. We then numerically evaluate the mixing time of the algorithm and compare how different values of $k$ impact the samples. Finally, we present numerical experiments showing that our model is able to reproduce more features of real networks than the Configuration Model, while being very different from the Correlated Configuration Model.

\section{Onion decomposition}

The $k$-core decomposition of a simple undirected graph is obtained by a pruning process whereby the lowest degree nodes are iteratively removed until no nodes of degree less than $k$ are in the resulting graph (the $k$-core). Applications of the $k$-core decomposition are often interested in the coreness of nodes~\cite{kitsak2010identification}, defined for each node as the largest integer $c$ such that the node is in the $c$-core but not in the $(c+1)$-core. While algorithms like that of Ref.~\cite{batagelj2011fast} are efficient in extracting the $k$-core decomposition, they throw away a lot of valuable information about how many iterations are necessary to remove a specific node.
The Onion Decomposition (OD)~\cite{hebert-dufresne_multi-scale_2016} preserves this information and can be obtained through a very simple modification of the pruning algorithm: alongside the core number $c$ of a node (or its \textit{coreness}), the iteration at which it is removed (the \textit{layer} $\ell$ to which it belongs) is kept in memory. By assigning a tuple $(c, \ell)$ to each node, the Onion Decomposition reveals how deep nodes are located within their $k$-core, thereby providing information about the \textit{internal} organization of each core.

While the $k$-core decomposition characterizes the meso- and macro- scales of a network (e.g. densely connected subgroups, core-periphery organization), the Onion Decomposition sheds light on the micro- and meso- scales.  For instance, it has been used to identify unlikely structures among the nodes of a $k$-core in real empirical networks~\cite{hebert-dufresne_multi-scale_2016} or to rank the position of nodes in the centrality structure of a network~\cite{young2019phase, mimar2022sampling, lu2024rk}. It also has been used in heuristics overcoming NP-hard optimization problems~\cite{garcia-perez2019mercator}, to parameterize dynamical systems of water distribution~\cite{zhou2023mesoscale} or organizational networks~\cite{hebert2023hierarchical}, and for the retrieval of the underlying filament structure of the cosmic web~\cite{bonnaire2020trex}.  

While it encodes interesting structures at multiple scales, the Onion Decomposition is still a \textit{local} measure in the sense that the layer of a node depends only on the layer of its immediate neighbors. This observation can be codified into the following \textit{local rules}:\\

\noindent\textbf{(Local rules)} Let $v$ be a node in graph $G$, with layer $\ell$ and coreness $c$. Then :
    \begin{enumerate}
        \item if $v$ is in the first layer of the core, it has exactly $c$ edges connected to nodes with layers $\ell' \geq \ell$.
        
        
        \item if $v$ is not in the first layer of the core, it has a) at least $c+1$ edges to nodes with layers $\ell' \geq \ell - 1$ and b) at most $c$ edges to nodes with layers $\ell' \geq \ell$.
    \end{enumerate}
These rules are the building blocks of the sampling algorithm described in the next section.

\section{Layered Configuration Model}

Since the Onion Decomposition contains information on the micro- meso- and macro- scales of a network and is easy to calculate, there is a very strong potential in using this measure as a constraint on random graph models. Many models that use the Onion Decomposition as a constraint have been used in the literature and have been shown to give more realistic representations of real world networks~\cite{hebert-dufresne_multi-scale_2016, allard_percolation_2019, hebertdufresne_2024_network}. In Ref.~\cite{hebert-dufresne_multi-scale_2016}, an ensemble with constraints on OD and layer-layer correlations is proposed, and these correlations are further constrained as to be between (degree, layer)-(degree, layer) tuples in Ref.~\cite{allard_percolation_2019}. These models are shown to be more justifiable than the CM in the case of sparse networks in Ref.~\cite{hebertdufresne_2024_network}. However, so far, there has been no rigorous analysis of the sampling algorithms for all of these ensembles. Furthermore, all these models have more constraints than simply the degree sequence and the Onion Decomposition, making it harder to analyze their sampling algorithms. We focus on sampling the simplest OD-based ensemble, which will in turn help to build robust algorithms for the other ensembles.

To follow the terminology of the Configuration Model, we dub \textit{Layered Configuration Model} (LCM) the ensemble containing all graphs with a fixed degree sequence $\{d_i\}_{i\in V}$ and a fixed Onion Decomposition $\{(c_i, \ell_i)\}_{i\in V}$. We refer to the LCM of a graph $G$ as the LCM spawned by the degree sequence and Onion Decomposition of $G$.

\subsection{Double-edge swap algorithm}

Consider a graph $G$ composed of $E$ edges and $V$ nodes. To sample from the LCM of $G$, we consider a MCMC algorithm based on edge swapping. A \textit{double edge swap} is an operation on a graph where two edges, say $(u,v)$ and $(x,y)$, are removed and replaced by two new edges, $(u,x);(v,y)$ or $(u,y);(v,x)$. Note that this operation does not affect the degree sequence of the graph, since all the nodes finish with the same number of neighbors. A Markov Chain composed of multiple consecutive edge swaps is the most common method to sample from the Configuration Model, since it has been shown to produce uniformly distributed samples~\cite{fosdick_configuring_2018, miklos_randomization_2004}. We consider the use of this same method when sampling from the LCM.

To respect the additional constraints of the Onion Decomposition, we use the \textit{Swap and Hold} (S\&H) method~\cite{artzyrandrup_2005_generating}. In this method, a swap is first proposed, and the obtained graph $G'$ is added to the sample only if the given constraints are respected. If $G'$ does not respect the constraints, then the swap is refused, and instead the previous graph $G$ is added to the sample. In the case of the LCM, we want to build an algorithm that looks, at every proposal, if a self-loop (edge from and to a single node) is created, if a multiedge (multiple edges between a single pair of nodes) is created, or if the Onion Decomposition is changed. The self-loop and multiedge constraints are necessary to ensure that a simple graph will be obtained. The procedure of the double-edge swap algorithm is detailed in algorithm~\ref{alg:LCM}, with $k=2$.

The main difficulty of the algorithm lies in the test to verify if the OD is changed by the proposed swap. A naive implementation would be to calculate the OD before and after the swap, check if there has been a change between the two measures, and refuse or accept the swap accordingly. However, since the calculation of OD is done in $\mathcal{O}(E\log V)$, this strategy is not computationally efficient and is not viable for large graphs. We instead provide a strategy built on the local rules of the OD, which allows an efficient $\mathcal{O}(1)$ verification.

We apply the following strategy: For each node involved in the swap, we look at its layer, its core, and the layers of its neighbors in the new graph $G'$. The test then makes a simple verification between these values to ensure that the local rules are still respected. The verification is based on the following observation: the local rules separate stubs (half-edges) in three categories: Those towards layers $\ell' > \ell$, those towards layer $\ell' = \ell - 1$ and those towards layers $\ell' < \ell - 1$. The test ensures that the number of stubs for each category falls in the accepted range given by the local rules.

Suppose we are testing the $2$-edge swap $(u,v);(x,y) \to (u,x);(v,y)$. If we first look at node $u$, the coreness $c_u$ and layer $\ell_u$ will indicate which rules we must look at: If $c(\ell_u) \neq c(\ell_u - 1)$ where $c(\ell)$ is the coreness of nodes in layer $\ell$, then we are in the first layer of core $c_u$. In this case, the rules ask that exactly $c_u$ stubs go towards nodes of layer $\ell_u$ or higher when the swap is made. In other words, if at the end of the swap there are still exactly $c_u$ stubs towards layers $\ell_u$ and higher, then we know the OD of node $u$ is not changed by the swap. Similarly, if $c(\ell_u) = c(\ell_u - 1)$, then we are not in the first layer of core $c_u$. In this case, we need to ensure that there are at least $c_u + 1$ nodes towards layers $\ell_u - 1$ and higher, and at most $c_u$ stubs towards layers $\ell_u$ and higher. If these rules are not respected, then we know that the layer of $u$ is changed during the swap. In this case, the OD is changed and the swap will be refused. Otherwise, the test is repeated on the three other nodes involved in the swap. Details are presented in algorithm~\ref{alg:OD_test}. 

\begin{algorithm}
\caption{OD conservation test}
\label{alg:OD_test}
\begin{algorithmic}
\REQUIRE Node $v$, its coreness $c_v$, its layer $\ell_v$, the layers of its new neighbors $\{\ell_i : (v, v_{i}) \in E(G') \}$ 
\ENSURE $\TRUE$ if the onion decomposition is changed by the $k$-edge swap, $\FALSE$ if it isn't
\IF{$\ell_v$ is the first layer of core $c_v$ (if $c(\ell_v) \neq c(\ell_v - 1)$)}
    \STATE $r \leftarrow$ number of $\ell_i \geq \ell_v$
    \IF{$r = c(v)$}
        \STATE \RETURN \ \FALSE
    \ELSE
        \STATE \RETURN \ \TRUE
    \ENDIF
\ELSIF{$\ell_v$ is not the first layer of core $c_v$}
    \STATE $r \leftarrow$ number of $\ell_i \geq \ell_v$
    \STATE $b \leftarrow$ number of $\ell_i = \ell_v - 1$
    \IF{$r\leq c(v)$ \textbf{and} $r+b \geq c(v) + 1$}
        \STATE \RETURN \ \FALSE
    \ELSE
        \STATE \RETURN \ \TRUE
    \ENDIF
\ENDIF
\end{algorithmic}   
\end{algorithm}

The MCMC algorithm can be considered as a random walk on a \textit{configuration graph}, where the nodes are the possible graphs $G_i$ from the LCM, and where there is a directed edge between $G_i$ and $G_j$ if there exists an edge swap that changes $G_i$ into $G_j$. Sampling the LCM uniformly requires this configuration graph to be 1) regular, 2) aperiodic, and 3) strongly connected~\cite{fosdick_configuring_2018}.

The advantage of using the S$\&$H method is that it guarantees the regularity and the aperiodicity of the configuration graph. Indeed, any swap from $G_i$ to $G_j$ that respects the constraints contributes 1 to the out-degree of graph $G_i$ and 1 to the in-degree of graph $G_j$. If a swap is refused, it is considered as a self-loop which will contribute 1 to both the in and out-degree of $G_i$. Since swaps are symmetric, the reverse move (from $G_j$ to $G_i$) will also be refused. Thus, every graph will have an in and out-degree of $2\binom{E}{2}$, and the configuration graph will always be regular. Furthermore, since a swap refusal corresponds to a self-loop (a cycle of length 1), the configuration graph will always be aperiodic \footnote{This will always happen for any graph with a node $u$ of degree 2 or more, since we can always take its two edges $(u,v)$ and $(u,x)$ and swap them. This will either end up with the same graph, or be refused because it creates a self-loop. In both cases, this is a self-loop in the configuration graph. If there is no node of degree $2$ or more, then all nodes are of degree $1$ and of layer $1$. In this case, the configuration graph of the LCM is equivalent to the one from the CM, which we know to be aperiodic~\cite{fosdick_configuring_2018}.}.

Since the chain is regular and aperiodic, we know that the sample will converge to the uniform distribution in the space it is able to explore. The only property that is therefore left to verify is the connectivity of the configuration graph.

\subsection{Counter-example of the connectivity}\label{sec:counter-examples}

To explore the connectivity of the configuration graph, we designed the following experiment: We first build a sample of the LCM of graph $G$ using the constrained double-edge swaps. Then, we build a sample of the CM of $G$ using regular double-edge swaps, and filter this sample to keep only the graphs with the same OD as $G$. Since we know the CM algorithm to be strongly connected, this method should, in principle, allow for the sampling of all graphs in the LCM of $G$. Then, comparing this filtered sample with the sample from the constrained swaps allows to see if the algorithm is disconnected. Since the size of both the LCM and the CM scale rapidly~\cite{hebertdufresne_2024_network}, this method is only viable for small graphs.

\begin{figure}
    \centering
    \includegraphics[width=1\linewidth]{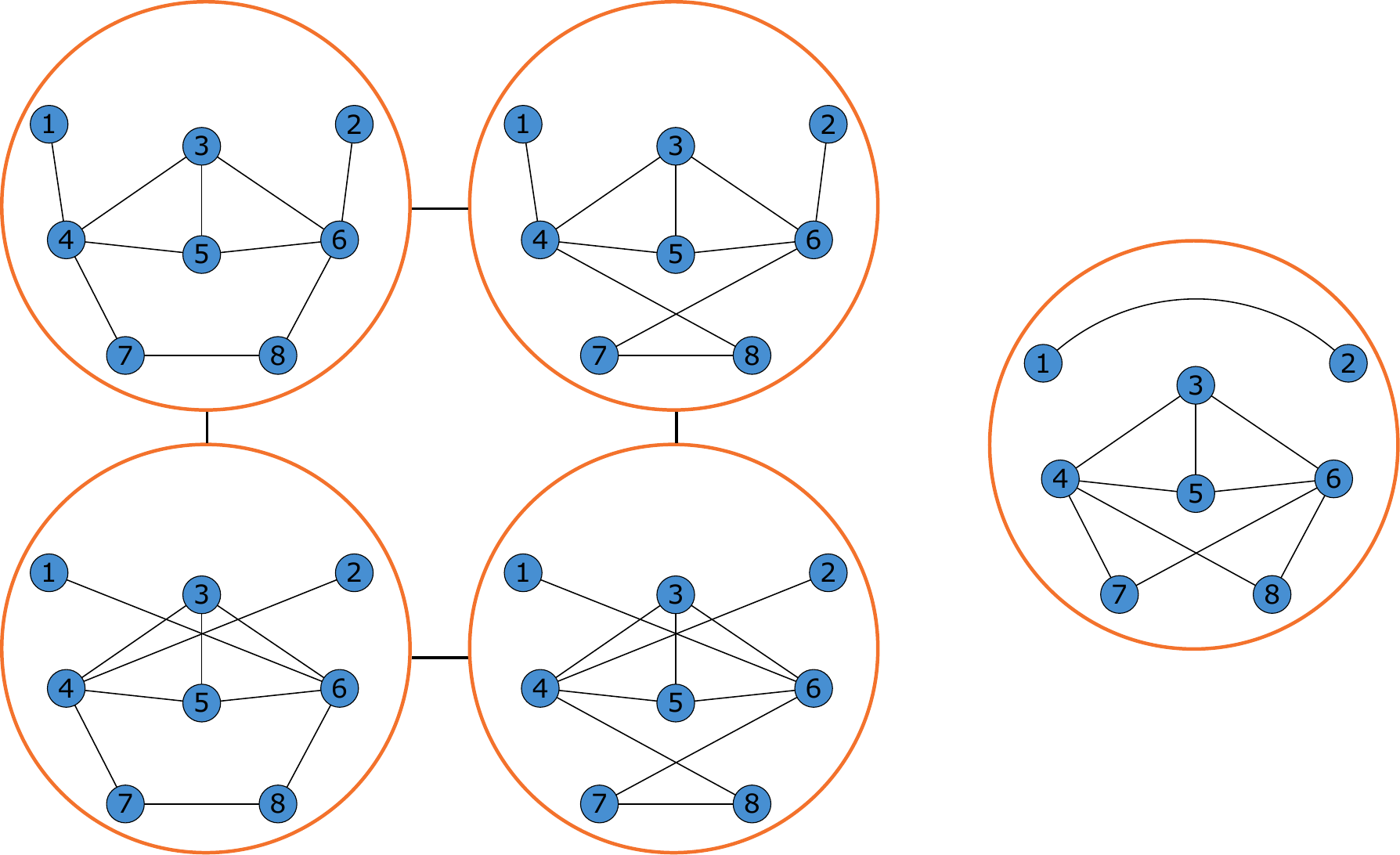}
    \caption{Complete configuration graph for the LCM of a graph with $N=8$ and $E=10$. An edge between two graphs indicates that a double-edge swap allows to pass from one graph to the other.}
    \label{fig:counter_example_connectivity}
\end{figure}

Using this method, we have found counter examples where double-edge swaps are not able to connect completely the configuration graph. One such example is shown in figure~\ref{fig:counter_example_connectivity}. Careful consideration shows that no combination of double-edge swaps can bring the graphs from the left component to the graph from the right component, even though they have the same OD and the same degree sequence.

\begin{figure}
    \centering
    \includegraphics[width=1\linewidth]{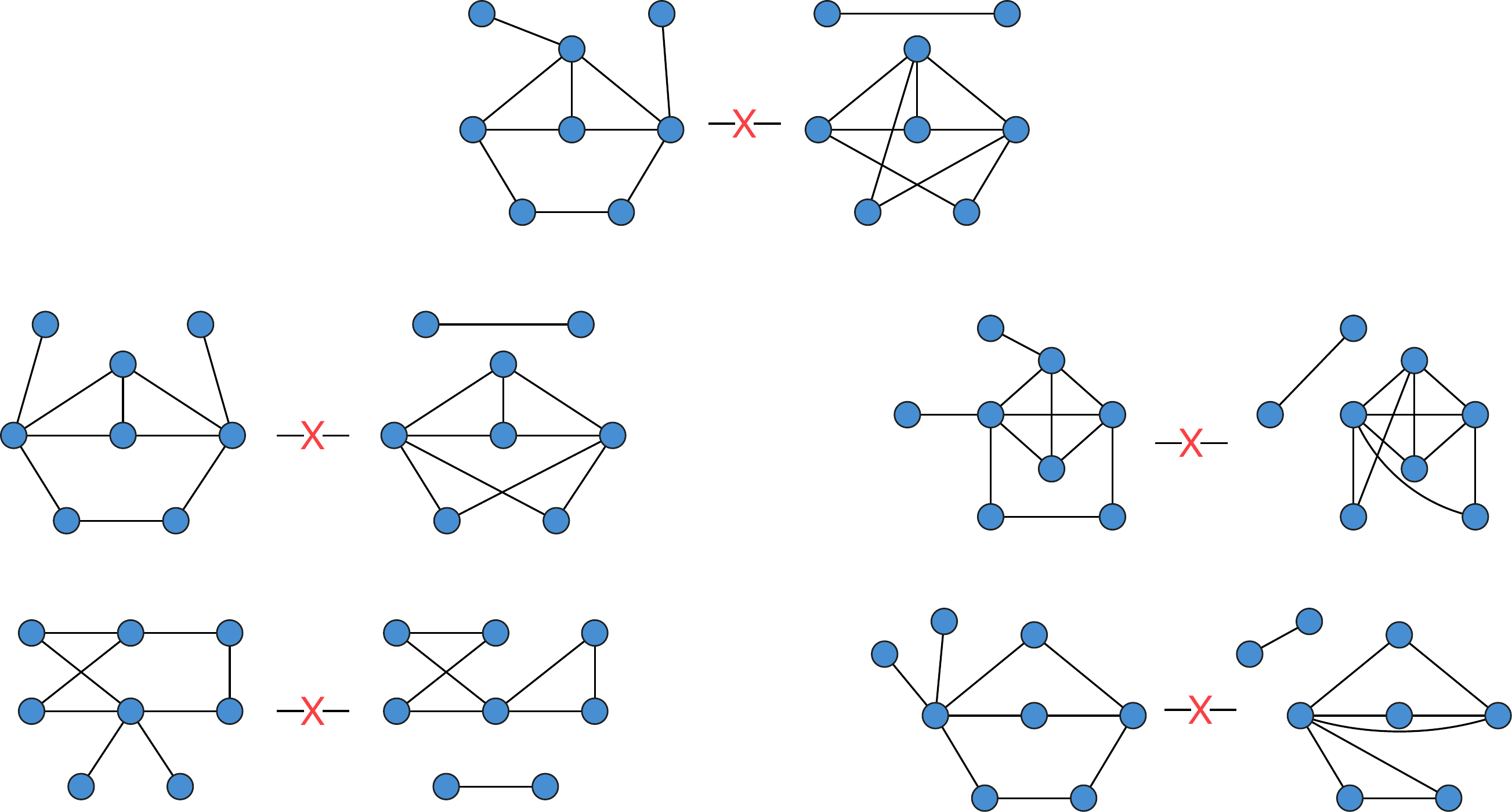}
    \caption{Some known counter-examples for which the LCM is not connected under double edge swaps. For each example, the graph on the left is taken from the connected component without the isolated edge, and the graph on the right is taken from the connected component with the isolated edge.}
    \label{fig:all_counter_examples}
\end{figure}

Figure~\ref{fig:all_counter_examples} shows $5$ counter-examples that were found using the filtering method. Interestingly, we have found that all these examples involve graphs with an isolated edge. In all the cases, double-edge swaps are not able to create this isolated edge without changing the OD, or without making a self-loop. Note that it is possible to create more counter-examples by adding isolate edges since it will still not be possible to swap these isolated edges with the principal component. 

It is possible to connect all these counter-examples if we allow the use of $3$-edge swaps, however. For example in figure~\ref{fig:counter_example_connectivity}, it is possible to swap edges $(1,2);(4,7);(6,8)$ into edges $(1,4);(2,6);(7,8)$ to connect the right graph to the left component. This hints to the possibility of sampling the complete LCM using higher-order edge swaps.

\subsection{$k$-edge swap algorithm}

We propose the more general algorithm~\ref{alg:LCM} to sample from the LCM using $k$-edge swaps. Much like the simpler double-edge swap, a $k$-edge swap consists of removing $k$ edges at random, say $\{u_i, v_i\}_{i=1}^k$, choosing a random permutation $\sigma$, and adding back the new edges $\{u_i, v_{\sigma(i)}\}_{i=1}^k$. 

With this method of random permutations, it is possible to pick $3$ edges and make only a $2$-edge swap, for example with $(u,v);(x,y);(w,z) \to (u,x);(v,y);(w,z)$. Thus, using $k$-edge swaps means that the algorithm also allows for $k' < k$ edge swaps. Therefore, increasing the value of $k$ will never reduce the number of graphs that can be reached. Furthermore, there is an upper bound to $k$ in which all graphs from the LCM can be reached from one another. Indeed, if graphs $G_i$ and $G_j$ differ by $x$ edges, an $x$-edge swap will connect graphs $G_i$ and $G_j$. Note that this upper bound remains generally unknown in practice.

The S\&H method guarantees a uniform sample over the graphs belonging to a same strongly connected component of the configuration graph. Furthermore, as we are allowing more complex swaps, more graphs become reachable and algorithm~\ref{alg:LCM} becomes increasingly likely to be strongly connected, and therefore to sample uniformly from the whole LCM.

However, as we increase $k$, proposed swaps become more likely to change the OD or to create self-loops or multiedge, and therefore to be rejected. Unfortunately, this downside cannot be completely avoided when using $k$-edge swaps and, as a consequence, will increase the mixing time of the algorithm, thereby highlighting the advantage of having efficient constraint verification algorithms (algorithm~\ref{alg:OD_test}) to perform the growing necessary number of swap proposals.

\begin{algorithm}
    \caption{$k$-edge swaps MCMC for the LCM}
    \label{alg:LCM}
    \begin{algorithmic}
    \REQUIRE initial simple graph $G_0$, number of sampled graphs $n$, order of edge swaps $k$
    \ENSURE sequence of graphs $G_i$
    \FOR{$i = 1$ to $n$}
        \STATE choose $k$ edges $\{u_i, v_i\}_{i=1}^k$ u.a.r. from $G_{i-1}$
        \STATE choose a random permutation $\sigma$ that sets the new edges $\{u_i, v_{\sigma(i)}\}_{i=1}^k$
        \STATE Change the chosen edges to the new edges to obtain graph $G'$
        \IF{$G'$ contains a self-loop or multiedge}
            \STATE set $G_i = G_{i-1}$
        \ELSIF{$G'$ does not have the OD of $G_{i-1}$ (see~\ref{alg:OD_test})}
            \STATE set $G_i = G_{i-1}$
        \ELSE
            \STATE set $G_i = G'$
        \ENDIF
    \ENDFOR
    \end{algorithmic}
\end{algorithm}

\section{Characterization of the algorithm}

Although allowing $k>2$ remedies the counter-examples listed in Sec.~\ref{sec:counter-examples} many unanswered questions remain: For which given degree sequence and Onion Decomposition are the configuration graphs disconnected? Is the isolated edge the only structure that makes the double edge swap disconnected? What part of the LCM are we missing when we use only double edge swaps, and how does this change when using higher values of $k$? How bad is the mixing time increased as we increase $k$?  By exploring samples with multiple values of $k$, we find some answers to these questions.

Many such constrained swap algorithms have been shown to be disconnected under double-edge swaps or have unknown connectivity \cite{preti_impossibility_2024, jamakovic_how_2015, orsini_quantifying_2015}. We believe that the the use of $k$-edge swaps, while less efficient, can still bring answers about the viability of sampling from the graph models.

\subsection{Mixing time}

We have seen earlier that the proposed edge swapping algorithm will eventually converge to the uniform distribution, but we have not yet answered the question of how long it takes for the chain to reach this distribution. This is an important aspect to consider, both to ensure that the algorithm can be used with a low computation time and to ensure that the obtained samples are not biased~\cite{dutta_sampling_2023}.

A heuristic numerical method has commonly been used to evaluate the convergence of the algorithm to its stationary state~\cite{gkantsidis_markov_2003, artzyrandrup_2005_generating, viger_efficient_2005, tabourier_generating_2011, dutta_sampling_2023}. This method uses the fact that most properties calculated on every graph from the sample will eventually reach a plateau given enough swapping, which corresponds to the equilibrium state of the Markov Chain. Notably, based on this heuristic method, Dutta et al.~\cite{dutta_sampling_2023} proposed guidelines to ensure a proper mixing of the samples for the CM.

\begin{figure}
    \centering
    \includegraphics[width=\linewidth]{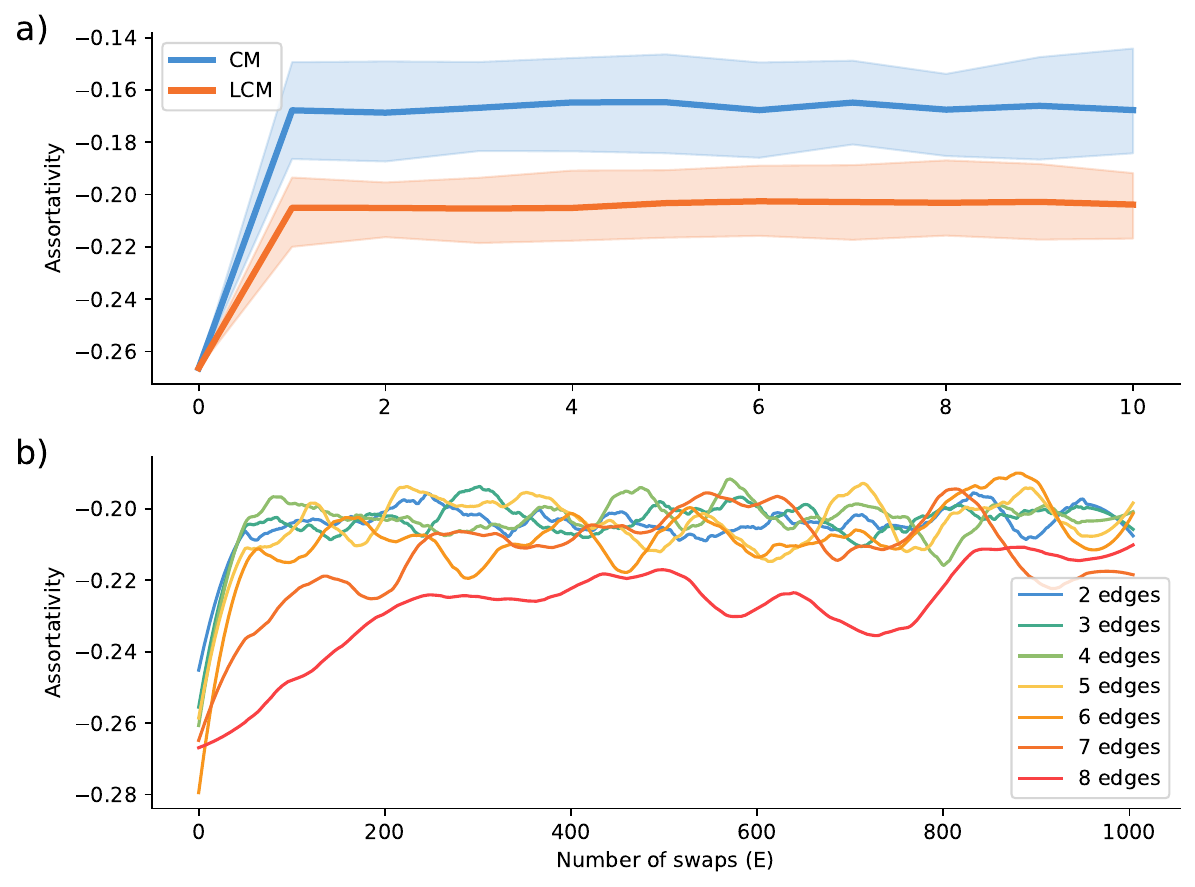}
    \caption{Network assortativity coefficient as a function of the number of attempted swaps of the Markov chain. The swaps have been performed on the Little Rock food web, $N=183$, $E=2494$~\cite{Martinez_littlerock_1991}. a) Comparison between $2$-edge swapping for the CM and the LCM.  A mean has been taken on 100 instances of the algorithms, and the 5th and 95th percentile over these instances are shown by the filled area.
    b) Comparison between different orders of $k$-edge swaps for the LCM. Curves have been smoothed with a Savitzky–Golay filter to help readability.}
    \label{fig:mixing_time}
\end{figure}

We use the same heuristic to compare the convergence of the CM algorithm and of our LCM algorithm. Figure~\ref{fig:mixing_time}a) shows the evolution of the assortativity coefficient when using double-edge swaps to shuffle a food web~\cite{newman2003mixing}. In both algorithms, we see that a plateau is reached in less than $1E$ attempted swaps. While we can expect our LCM algorithm to refuse more swaps and thus to have a slower mixing time, we can see that in practice, this does not significantly affect the mixing time compared to the CM algorithm.

So far, we have not found any network in which the plateau of the LCM is reached in a significantly higher number of swaps than for the CM. This means that in general, we can expect the S\&H algorithm to rapidly reach a uniform distribution, and we can expect the graphs from our samples to be statistically independent~\cite{dutta_sampling_2023}. As a safety measure, the samples that will be used in the rest of the article will all be separated by $400E$ attempted swaps. We expect this number to be sufficient for our purposes.

In Fig.~\ref{fig:mixing_time}b), we show the assortativity coefficient of a single chain for multiple values of $k$ when shuffling the LCM. We observe a difference in the convergence to the plateau, with higher-order edge swaps being slower, and with the $8$-edge swapping not having reached it yet after a $1000E$ swaps. This is expected, since there is an increase in the number of refused swaps with an increase in the value of $k$. This experiment shows that there is a need to increase the number of swaps if we need to use higher values of $k$, and is a limit to the current algorithm.

\subsection{Topological features while varying $k$}

Since higher values of $k$ lead to more swaps being refused, we want to use the algorithm with a minimal value of $k$. In fact, it would be ideal to use the $2$-edge swap algorithm in terms of convergence, even though we have shown that it is not always connected. In this section, we will compare samples obtained from algorithms with different values of $k$ and show that the case of $k=2$ can still be used in practical cases.

\begin{figure}
    \centering
    \includegraphics[width=\linewidth]{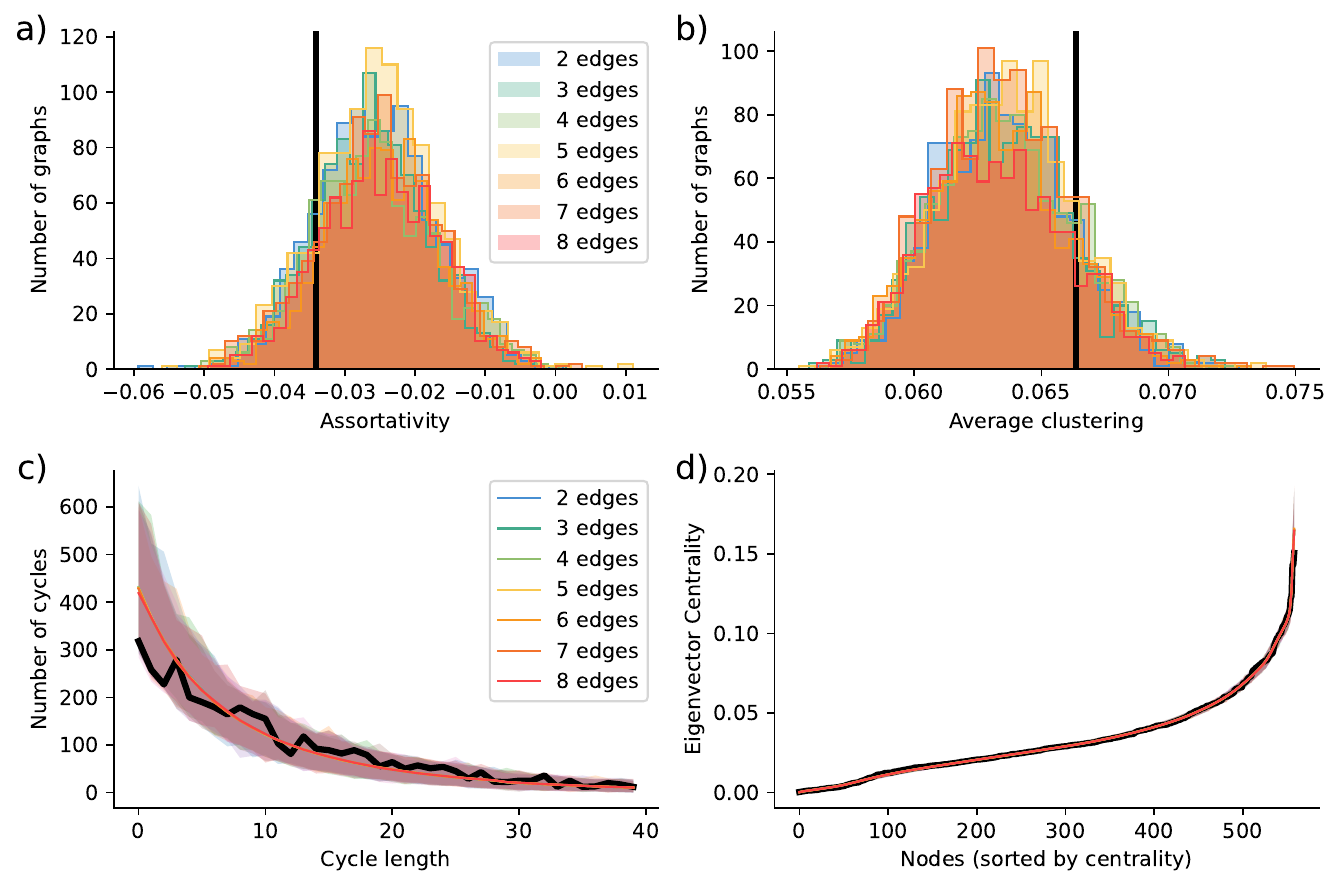}
    \caption{Comparison of topological properties of graph samples for different values of $k$. All samples are from the LCM of the complete \textit{C. elegans} connectome, $N=575$, $E=5306$,~\cite{Cook2019_male_chemical_corrected}. a) Degree assortativity b) Average clustering coefficient c) Cycle lengths d) Eigenvector centrality. The solid black lines show the properties of the original network, while the filled area in c) and d) show the minimal and maximal values obtained for each sample.}
    \label{fig:k_edges_properties}
\end{figure}

First, from Fig.\ref{fig:mixing_time}b, we can observe that the chains all converge to the same plateau of assortativity coefficient (except in the case of the $8$-edge swap, which has not fully converged yet). This is is hinting to the fact that the samples we obtain are similar. In Fig.~\ref{fig:k_edges_properties}, we take this analysis further by looking at different graph measures. For all chosen measure on the samples, we obtain similar distributions for values of $k$ ranging from $2$ to $8$. Thus, we observe that using more complex moves to sample from the LCM does not add any new information to the obtained samples. In other words, the graphs that were not part of the largest strongly connected component in the configuration graphs did not appear to be significantly different to the ones in it. While we show the results for a specific graph here, the conclusion is unanimous in all tested empirical networks.

\begin{figure}
    \centering
    \includegraphics[width=\linewidth]{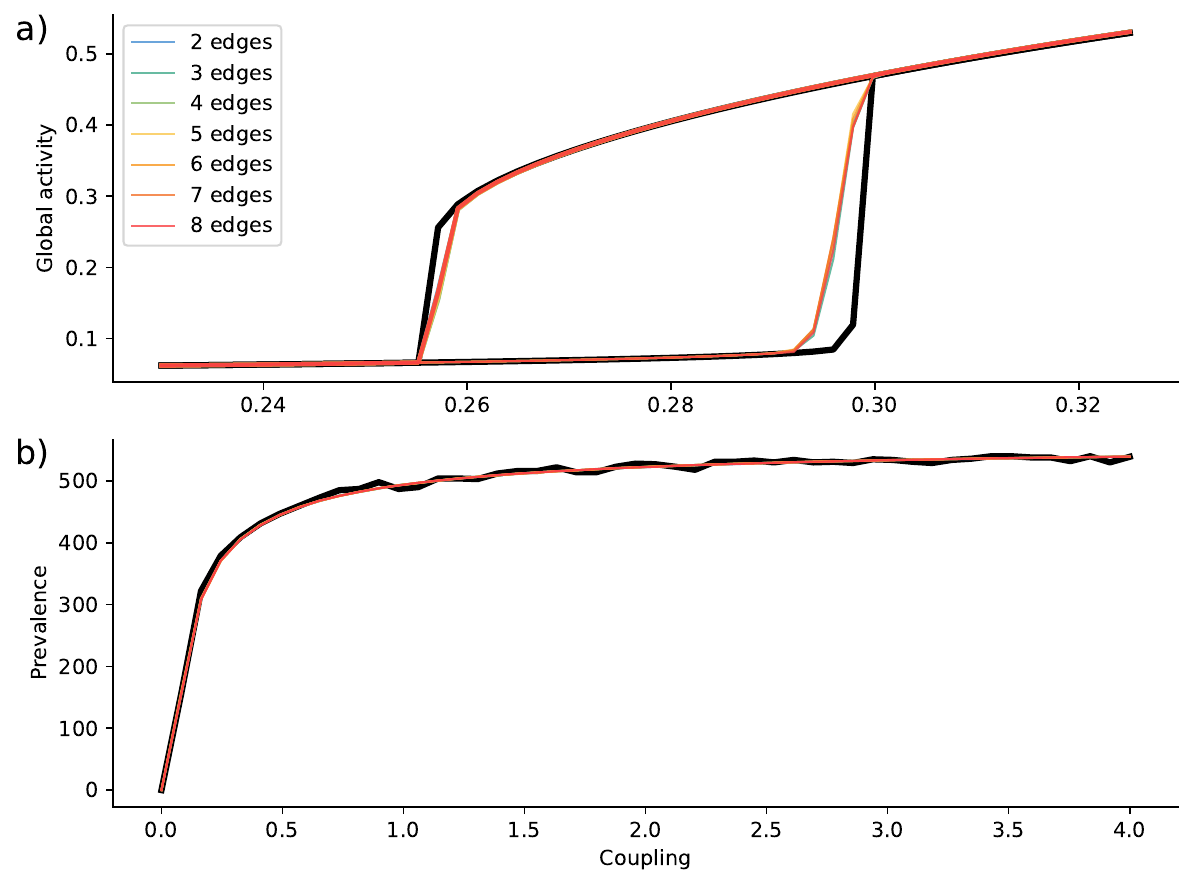}
    \caption{Bifurcation diagrams of dynamics over samples of the LCM for different values of $k$. The LCMs are based on the complete \textit{c. elegans} connectome. a) Wilson-Cowan dynamics b) SIS dynamics. In all cases, a mean is taken over the whole sample. The solid black line represents the behaviour of the dynamics over the real network.}
    \label{fig:k_edges_dynamics}
\end{figure}

In addition to looking at the structural properties of the samples, we compare the variation in their functional properties by comparing the behavior of dynamics evolving over them. We observe the behaviour of both a deterministic Wilson-Cowan (WC)~\cite{Thibeault2024} and a stochastic Susceptible-Infected-Susceptible (SIS)~\cite{Miller2019, st-onge2019efficient}. Fig.~\ref{fig:k_edges_dynamics}a) shows the hysteresis in the global activity of the WC dynamics, and Fig.~\ref{fig:k_edges_dynamics}b) shows the prevalence of the SIS. In both cases, we observe very little difference in the behaviour of the dynamics on the different samples. In other words, how the sample is mixed does not influence the functional properties of the samples, which further supports the idea that more complex swapping is not necessary to sample from these LCMs.

In light of these experiments, we can confidently say that we obtain similar samples when using $k$-edge swaps with $k$ ranging from $2$ to $8$. There are three ways to interpret these results: First, it can mean that the $2$-edge swap algorithm is not disconnected for the graphs that were tested. In this case, we have shown that not all LCMs are disconnected through double edge swaps, and that the simplest (and most efficient) version of the algorithm can be used in practical cases. Second, if the double edge swap algorithm is indeed unable to sample some graphs and the other algorithms are able to, then this means that for most purposes, these missed graphs are not necessary to obtain a representative sample of the LCM. Finally, we also need to consider the possibility that $8$-edge swaps are not enough to recover the disconnected part of the algorithm, and that all samples are incomplete. In all cases, we recommend careful consideration when drawing samples of the LCM, and propose to always compare different values of $k$ to strengthen confidence in results.

For the next section, results on the LCM are taken with the double-edge swap algorithm only.

\subsection{Comparison to CM and CCM}

Next, we compare how the LCM fares when compared to both the Configuration Model and the Correlated Configuration Model. We will see that enforcing the Onion Decomposition affects various topological and functional properties of the samples, and that the samples are often more representative of real networks than their more common counterparts.

\begin{figure}
    \centering
    \includegraphics[width=\linewidth]{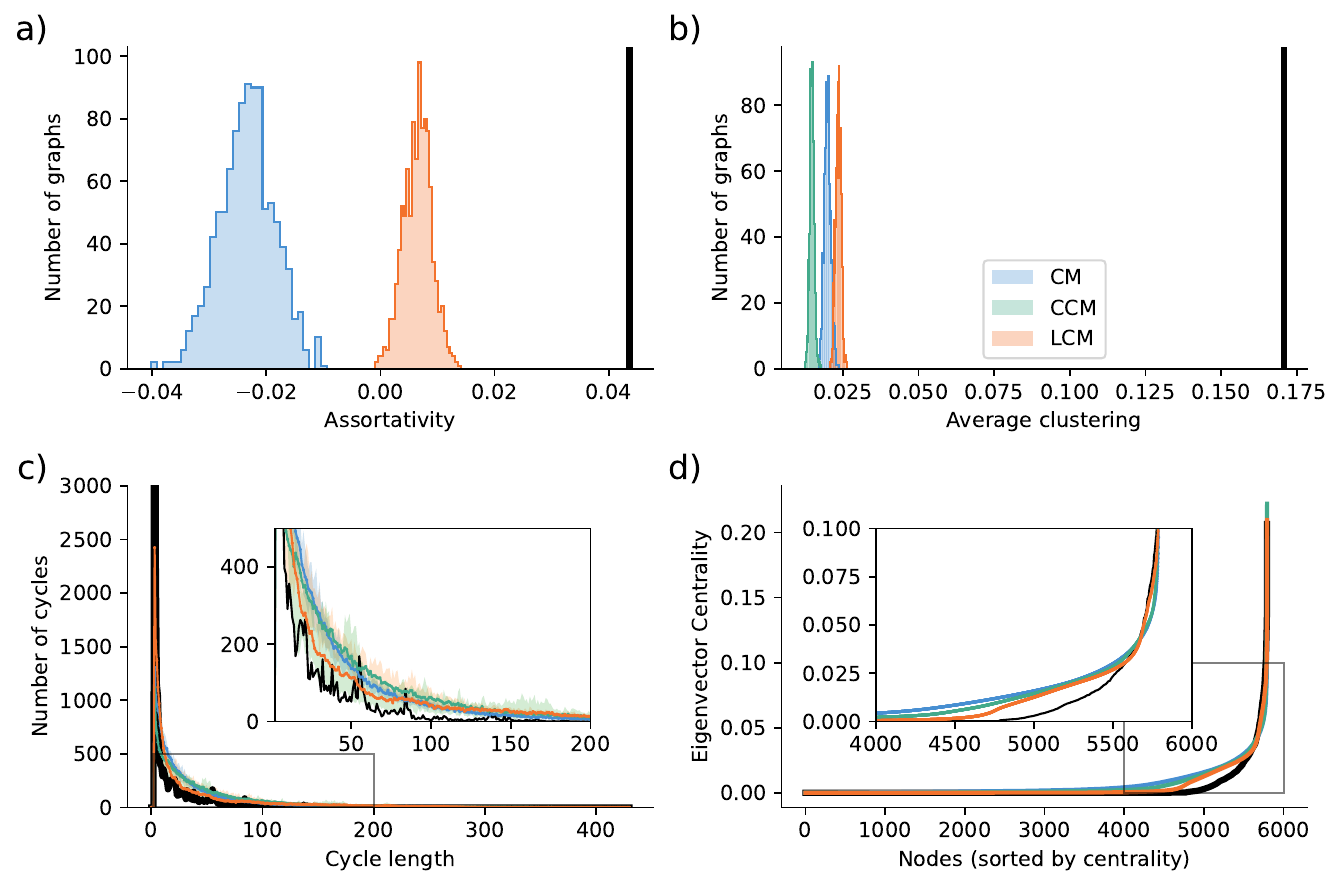}
    \caption{Comparison of the topological properties of samples from the CM, the CCM and the LCM. Samples and original measures are taken on a within-organization Facebook friendship network $N=5793$, $E=45266$~\cite{Fire2016_facebookL1}. The solid black line indicates values on the real network, while the filled area are the minimal and maximal values of the samples. In the case of the degree assortativity, the CCM distribution is not shown because it is kept exactly by definition.}
    \label{fig:ensemble_properties}
\end{figure}

In most empirical networks, we find that the graphs picked from the LCM are much closer in terms of properties than in the CM. Figure~\ref{fig:ensemble_properties} shows such an example on a Facebook friendship network. In this case, we see that the LCM keeps important structural properties of the original network, notably its assortativity, its clustering (although only on a small scale), the cycles and the centrality of the nodes. In this particular case, the graphs sampled from the LCM are also closer to the real network than the ones from the CCM, meaning that the OD is keeping more structure than the degree-degree correlations. This is not always the case however, depending on the OD or joint degree matrix of the observed network.

The ability of the OD to preserve larger structural features (for example the cycles or the centrality) comes from the fact that the pruning process to calculate it involves the complete neighborhood of every node. This is in opposition to both degree sequence and degree-degree correlations, which mainly enforce local structure. In other words, even if the LCM is sampled using only local information in the local rules, it is able to encompass more macro-scale information than the other models.

\begin{figure}
    \centering
    \includegraphics[width=\linewidth]{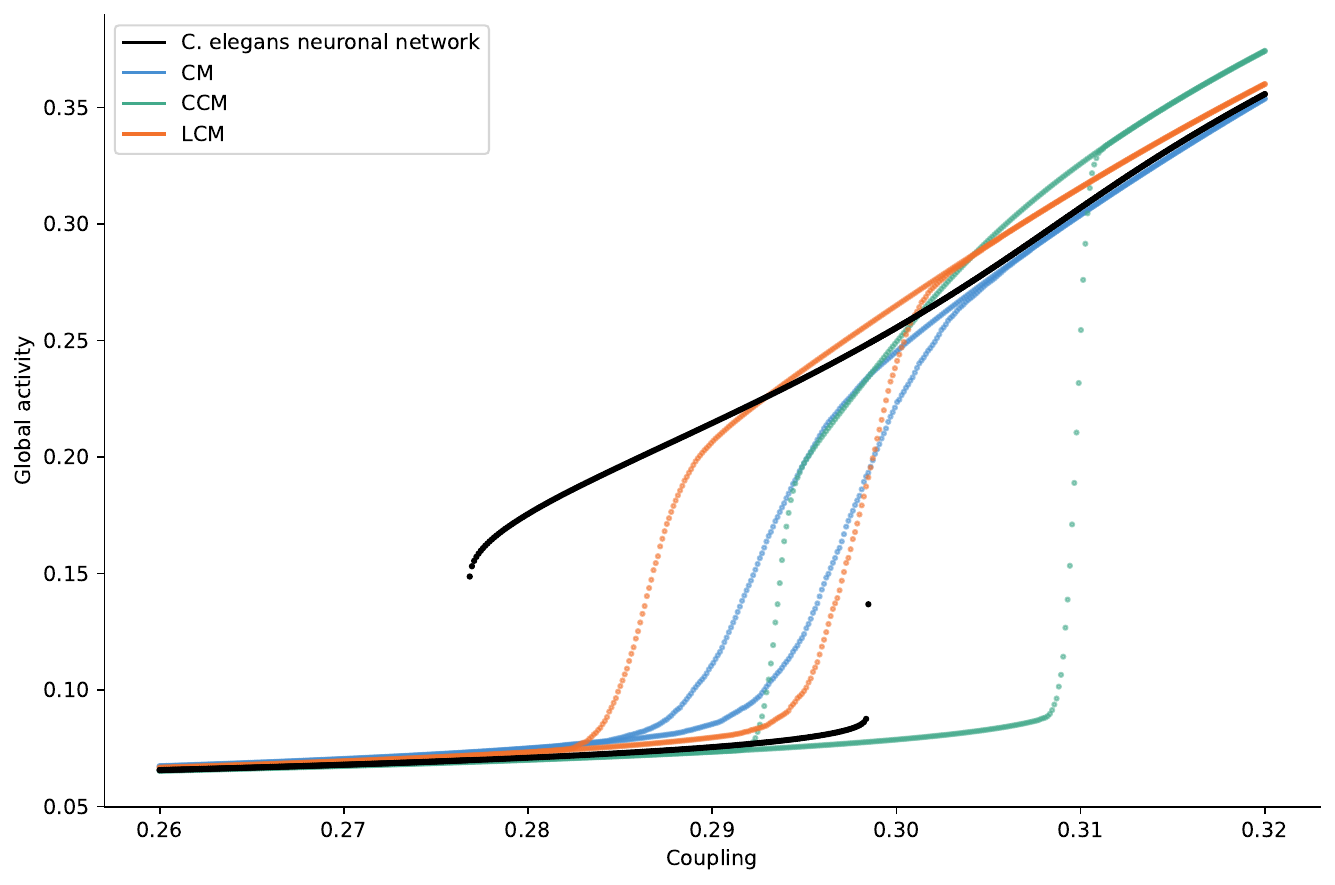}
    \caption{Mean global activity of a Wilson-Cowan dynamics on samples from the CM, CCM and LCM for different values of coupling. The black solid line represents the dynamics on the original network of \textit{C. elegans} neural connections, $N=297$, $E=2359$,~\cite{White_celegansneural}.}
    \label{fig:WC}
\end{figure}

\begin{figure}
    \centering
    \includegraphics[width=\linewidth]{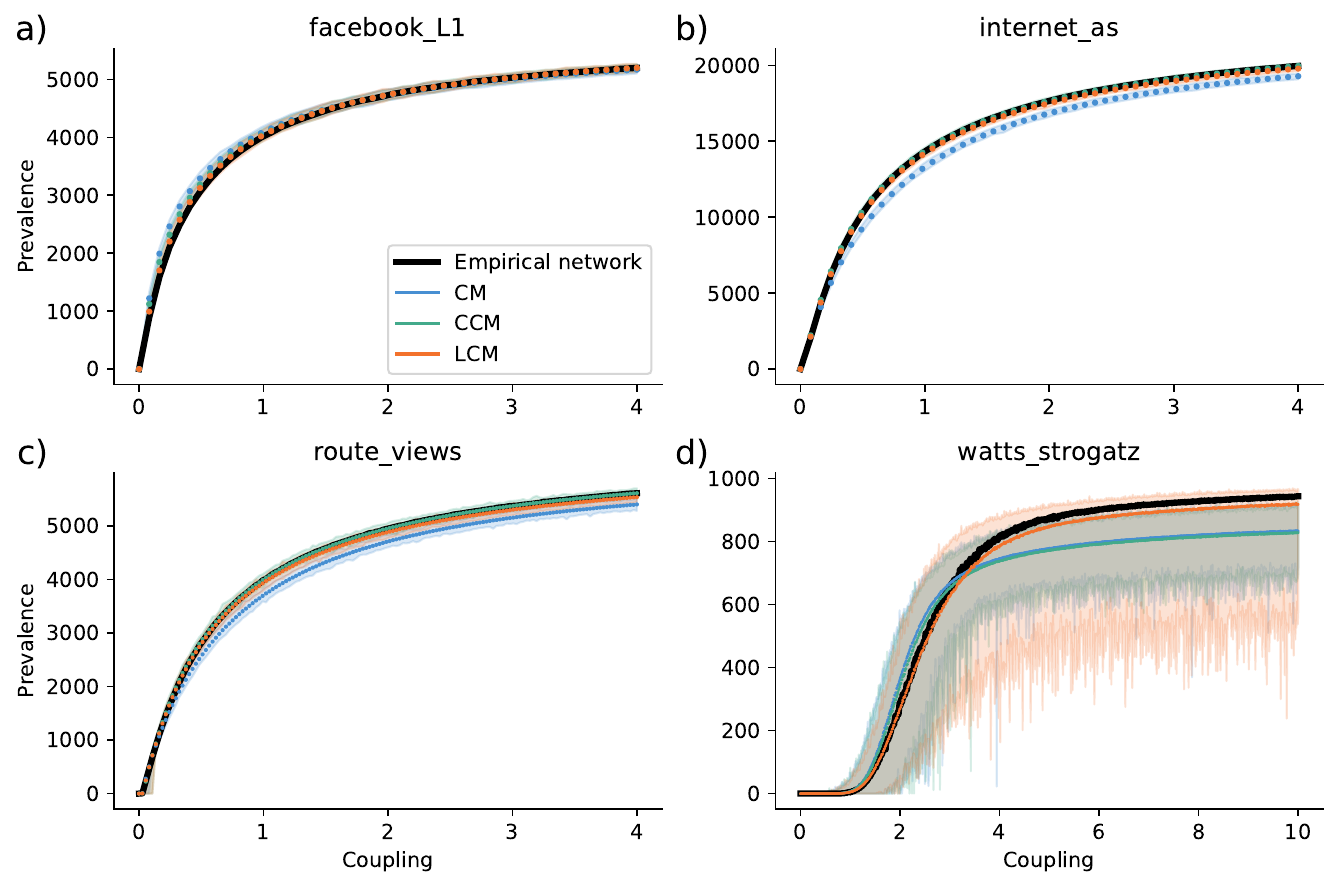}
    \caption{SIS dynamics prevalence for different values of coupling (infection probability). The solid black line represents the mean prevalence over 100 simulations of the SIS on the real network. The points are the mean taken over single simulations on $1000$ graphs from the CM, the CCM and the LCM. The filled area gives the maximal and minimal values. a) Facebook friendships $N=5793$, $E=45266$~\cite{Fire2016_facebookL1}. b) Internet Autonomous Systems $N=22963$, $E=48436$~\cite{internet_as}. c) BGP AS Internet traffic $N=6474$, $E=13895$~\cite{route_views_bgp_data}. d) Watts-Strogatz random network with rewiring parameter $0.1$, $N=1000$, $E=1000$~\cite{Watts1998_celegansneural}.}
    \label{fig:SIS_ensemble_comparison}
\end{figure}

Fig.~\ref{fig:WC} shows the bifurcation diagram of the Wilson-Cowan dynamics on the \textit{c. elegans} connectome~\cite{Watts1998_celegansneural} and on graphs sampled from the three models. We see that in this case, the LCM is better at representing the hysteresis found in the dynamics. In the same way, Fig.~\ref{fig:SIS_ensemble_comparison} shows how the LCM fares when trying to reproduce the stochastic SIS dynamics. In all cases, we observe that the LCM is much more representative of the dynamics of the real network than the CM. When comparing to the CCM, the difference is less evident, with the CCM often being more accurate. This is to be expected, as degree-degree correlations are highly influential in the context of a spreading dynamics \cite{Gupta1989NetworksOS, VanMieghem2010, Mata_2014}. The experiments here show that the LCM is also able to enforce some degree-degree correlations, which explains its success when reproducing SIS dynamics.

In the case of the Watts-Strogatz random network in Fig.~\ref{fig:SIS_ensemble_comparison}d), we observe that the LCM fares significantly better. This is explained by the fact that there is close to no variance in the degree sequence of the graph, and therefore the LCM captures more structure than pure degree-based models like the CM and the CCM.

\section{Conclusion}

Our first contribution in this paper is the definition of the simplest random graph model to account for the Onion Decomposition, the Layered Configuration Model, which had not been proposed or explored yet in the literature. We proposed a set of $k$-edge swapping algorithms to obtain uniform samples from this model, and have shown that the lowest values of $k$ have a good mixing time. While we have no definitive answer about the connectivity of these edge swapping algorithms, we have provided a numerical experiment to test how changing the value of $k$ affect the connectivity of the space. Our results point to the fact that $2$-edge swaps are sufficient in most practical cases.

Comparing the Layered Configuration Model with the Configuration Model, we have shown that adding a simple constraint on the Onion Decomposition is extremely informative on many empirical networks. This highlights the importance of considering meso-scale structures in the development of null models. We believe that in this matter, the Onion Decomposition is an unexpectedly strong statistic since it encompasses macro- and meso-scale structures while being easy to sample from since it is based on degrees and on local neighborhoods.

Finally, the comparison of the Layered Configuration Model with the Correlated Configuration Model shows that there is still room for improvement if we want null models that reproduce more features of empirical networks. Notably, the importance of degree-degree correlations in the context of the SIS dynamics leads to the natural generalization of adding correlation, either degree-degree, layer-layer, or (degree,layer)-(degree,layer), as constraints on the Layered Configuration Model. This could be done either on average with the method of~\cite{newman_2002_assortative}, or with exact constraints as has been proposed in Refs.~\cite{hebert-dufresne_multi-scale_2016, allard_percolation_2019, hebertdufresne_2024_network}. While these more constrained ensembles have already been proposed and explored, there is still a need to develop rigorous and efficient sampling algorithms for these models. We believe that the proposed work here provides a strong basis upon which to build richer null models based on local connection rules that encode non-local network properties.

\begin{acknowledgments}
  This work was supported by the Sentinelle Nord program of Universit\'e Laval funded by the Canada First Research Excellence Fund (FT, AA), the Vermont Complex Systems Center (LHD), and the Natural Sciences and Engineering Research Council of Canada (AA).
\end{acknowledgments}





\begin{thebibliography}{67}%
\makeatletter
\providecommand \@ifxundefined [1]{%
 \@ifx{#1\undefined}
}%
\providecommand \@ifnum [1]{%
 \ifnum #1\expandafter \@firstoftwo
 \else \expandafter \@secondoftwo
 \fi
}%
\providecommand \@ifx [1]{%
 \ifx #1\expandafter \@firstoftwo
 \else \expandafter \@secondoftwo
 \fi
}%
\providecommand \natexlab [1]{#1}%
\providecommand \enquote  [1]{``#1''}%
\providecommand \bibnamefont  [1]{#1}%
\providecommand \bibfnamefont [1]{#1}%
\providecommand \citenamefont [1]{#1}%
\providecommand \href@noop [0]{\@secondoftwo}%
\providecommand \href [0]{\begingroup \@sanitize@url \@href}%
\providecommand \@href[1]{\@@startlink{#1}\@@href}%
\providecommand \@@href[1]{\endgroup#1\@@endlink}%
\providecommand \@sanitize@url [0]{\catcode `\\12\catcode `\$12\catcode
  `\&12\catcode `\#12\catcode `\^12\catcode `\_12\catcode `\%12\relax}%
\providecommand \@@startlink[1]{}%
\providecommand \@@endlink[0]{}%
\providecommand \url  [0]{\begingroup\@sanitize@url \@url }%
\providecommand \@url [1]{\endgroup\@href {#1}{\urlprefix }}%
\providecommand \urlprefix  [0]{URL }%
\providecommand \Eprint [0]{\href }%
\providecommand \doibase [0]{https://doi.org/}%
\providecommand \selectlanguage [0]{\@gobble}%
\providecommand \bibinfo  [0]{\@secondoftwo}%
\providecommand \bibfield  [0]{\@secondoftwo}%
\providecommand \translation [1]{[#1]}%
\providecommand \BibitemOpen [0]{}%
\providecommand \bibitemStop [0]{}%
\providecommand \bibitemNoStop [0]{.\EOS\space}%
\providecommand \EOS [0]{\spacefactor3000\relax}%
\providecommand \BibitemShut  [1]{\csname bibitem#1\endcsname}%
\let\auto@bib@innerbib\@empty
\bibitem [{\citenamefont {Fosdick}\ \emph {et~al.}(2018)\citenamefont
  {Fosdick}, \citenamefont {Larremore}, \citenamefont {Nishimura},\ and\
  \citenamefont {Ugander}}]{fosdick_configuring_2018}%
  \BibitemOpen
  \bibfield  {author} {\bibinfo {author} {\bibfnamefont {B.~K.}\ \bibnamefont
  {Fosdick}}, \bibinfo {author} {\bibfnamefont {D.~B.}\ \bibnamefont
  {Larremore}}, \bibinfo {author} {\bibfnamefont {J.}~\bibnamefont
  {Nishimura}},\ and\ \bibinfo {author} {\bibfnamefont {J.}~\bibnamefont
  {Ugander}},\ }\bibfield  {title} {\bibinfo {title} {Configuring {Random}
  {Graph} {Models} with {Fixed} {Degree} {Sequences}},\ }\href
  {https://doi.org/10.1137/16M1087175} {\bibfield  {journal} {\bibinfo
  {journal} {SIAM Rev.}\ }\textbf {\bibinfo {volume} {60}},\ \bibinfo {pages}
  {315} (\bibinfo {year} {2018})}\BibitemShut {NoStop}%
\bibitem [{\citenamefont {Carstens}\ and\ \citenamefont
  {Horadam}(2017)}]{carstens_switching_2017}%
  \BibitemOpen
  \bibfield  {author} {\bibinfo {author} {\bibfnamefont {C.~J.}\ \bibnamefont
  {Carstens}}\ and\ \bibinfo {author} {\bibfnamefont {K.~J.}\ \bibnamefont
  {Horadam}},\ }\bibfield  {title} {\bibinfo {title} {Switching edges to
  randomize networks: what goes wrong and how to fix it},\ }\href
  {https://doi.org/10.1093/comnet/cnw027} {\bibfield  {journal} {\bibinfo
  {journal} {J. Complex Netw.}\ }\textbf {\bibinfo {volume} {5}},\ \bibinfo
  {pages} {337} (\bibinfo {year} {2017})}\BibitemShut {NoStop}%
\bibitem [{\citenamefont {Kim}\ \emph {et~al.}(2009)\citenamefont {Kim},
  \citenamefont {Toroczkai}, \citenamefont {Erdős}, \citenamefont {Miklós},\
  and\ \citenamefont {Székely}}]{kim_degree-based_2009}%
  \BibitemOpen
  \bibfield  {author} {\bibinfo {author} {\bibfnamefont {H.}~\bibnamefont
  {Kim}}, \bibinfo {author} {\bibfnamefont {Z.}~\bibnamefont {Toroczkai}},
  \bibinfo {author} {\bibfnamefont {P.~L.}\ \bibnamefont {Erdős}}, \bibinfo
  {author} {\bibfnamefont {I.}~\bibnamefont {Miklós}},\ and\ \bibinfo {author}
  {\bibfnamefont {L.~A.}\ \bibnamefont {Székely}},\ }\bibfield  {title}
  {\bibinfo {title} {Degree-based graph construction},\ }\href
  {https://doi.org/10.1088/1751-8113/42/39/392001} {\bibfield  {journal}
  {\bibinfo  {journal} {J. Phys. A}\ }\textbf {\bibinfo {volume} {42}},\
  \bibinfo {pages} {392001} (\bibinfo {year} {2009})}\BibitemShut {NoStop}%
\bibitem [{\citenamefont {Genio}\ \emph {et~al.}(2010)\citenamefont {Genio},
  \citenamefont {Kim}, \citenamefont {Toroczkai},\ and\ \citenamefont
  {Bassler}}]{genio_efficient_2010}%
  \BibitemOpen
  \bibfield  {author} {\bibinfo {author} {\bibfnamefont {C.~I.~D.}\
  \bibnamefont {Genio}}, \bibinfo {author} {\bibfnamefont {H.}~\bibnamefont
  {Kim}}, \bibinfo {author} {\bibfnamefont {Z.}~\bibnamefont {Toroczkai}},\
  and\ \bibinfo {author} {\bibfnamefont {K.~E.}\ \bibnamefont {Bassler}},\
  }\bibfield  {title} {\bibinfo {title} {Efficient and {Exact} {Sampling} of
  {Simple} {Graphs} with {Given} {Arbitrary} {Degree} {Sequence}},\ }\href
  {https://doi.org/10.1371/journal.pone.0010012} {\bibfield  {journal}
  {\bibinfo  {journal} {PLOS ONE}\ }\textbf {\bibinfo {volume} {5}},\ \bibinfo
  {pages} {e10012} (\bibinfo {year} {2010})}\BibitemShut {NoStop}%
\bibitem [{\citenamefont {Blitzstein}\ and\ \citenamefont
  {Diaconis}(2011)}]{diaconis_sequential_2011}%
  \BibitemOpen
  \bibfield  {author} {\bibinfo {author} {\bibfnamefont {J.}~\bibnamefont
  {Blitzstein}}\ and\ \bibinfo {author} {\bibfnamefont {P.}~\bibnamefont
  {Diaconis}},\ }\bibfield  {title} {\bibinfo {title} {A {{Sequential
  Importance Sampling Algorithm}} for {{Generating Random Graphs}} with
  {{Prescribed Degrees}}},\ }\href
  {https://doi.org/10.1080/15427951.2010.557277} {\bibfield  {journal}
  {\bibinfo  {journal} {Internet Math.}\ }\textbf {\bibinfo {volume} {6}},\
  \bibinfo {pages} {489} (\bibinfo {year} {2011})}\BibitemShut {NoStop}%
\bibitem [{\citenamefont {Kannan}\ \emph {et~al.}(1999)\citenamefont {Kannan},
  \citenamefont {Tetali},\ and\ \citenamefont {Vempala}}]{Kannan_1999_simple}%
  \BibitemOpen
  \bibfield  {author} {\bibinfo {author} {\bibfnamefont {R.}~\bibnamefont
  {Kannan}}, \bibinfo {author} {\bibfnamefont {P.}~\bibnamefont {Tetali}},\
  and\ \bibinfo {author} {\bibfnamefont {S.}~\bibnamefont {Vempala}},\
  }\bibfield  {title} {\bibinfo {title} {Simple markov-chain algorithms for
  generating bipartite graphs and tournaments},\ }\href
  {https://doi.org/https://doi.org/10.1002/(SICI)1098-2418(199907)14:4<293::AID-RSA1>3.0.CO;2-G}
  {\bibfield  {journal} {\bibinfo  {journal} {Random Struct. Algorithms}\
  }\textbf {\bibinfo {volume} {14}},\ \bibinfo {pages} {293} (\bibinfo {year}
  {1999})}\BibitemShut {NoStop}%
\bibitem [{\citenamefont {Cooper}\ \emph {et~al.}(2007)\citenamefont {Cooper},
  \citenamefont {Dyer},\ and\ \citenamefont
  {Greenhill}}]{cooper_sampling_2007}%
  \BibitemOpen
  \bibfield  {author} {\bibinfo {author} {\bibfnamefont {C.}~\bibnamefont
  {Cooper}}, \bibinfo {author} {\bibfnamefont {M.}~\bibnamefont {Dyer}},\ and\
  \bibinfo {author} {\bibfnamefont {C.}~\bibnamefont {Greenhill}},\ }\bibfield
  {title} {\bibinfo {title} {Sampling {Regular} {Graphs} and a {Peer}-to-{Peer}
  {Network}},\ }\href {https://doi.org/10.1017/S0963548306007978} {\bibfield
  {journal} {\bibinfo  {journal} {Comb. Probab. Comput.}\ }\textbf {\bibinfo
  {volume} {16}},\ \bibinfo {pages} {557} (\bibinfo {year} {2007})}\BibitemShut
  {NoStop}%
\bibitem [{\citenamefont {Cooper}\ \emph {et~al.}(2012)\citenamefont {Cooper},
  \citenamefont {Dyer},\ and\ \citenamefont
  {Greenhill}}]{cooper2012corrigendumsamplingregulargraphs}%
  \BibitemOpen
  \bibfield  {author} {\bibinfo {author} {\bibfnamefont {C.}~\bibnamefont
  {Cooper}}, \bibinfo {author} {\bibfnamefont {M.}~\bibnamefont {Dyer}},\ and\
  \bibinfo {author} {\bibfnamefont {C.}~\bibnamefont {Greenhill}},\ }\href
  {http://arxiv.org/abs/1203.6111} {\emph {\bibinfo {title} {Corrigendum:
  {{Sampling}} regular graphs and a peer-to-peer network}}},\ \bibinfo {type}
  {Preprint}\ \bibinfo {number} {arXiv:1203.6111}\ (\bibinfo {year}
  {2012})\BibitemShut {NoStop}%
\bibitem [{\citenamefont {Erdős}\ \emph {et~al.}(2018)\citenamefont {Erdős},
  \citenamefont {Miklós},\ and\ \citenamefont {Toroczkai}}]{erdos_new_2018}%
  \BibitemOpen
  \bibfield  {author} {\bibinfo {author} {\bibfnamefont {P.~L.}\ \bibnamefont
  {Erdős}}, \bibinfo {author} {\bibfnamefont {I.}~\bibnamefont {Miklós}},\
  and\ \bibinfo {author} {\bibfnamefont {Z.}~\bibnamefont {Toroczkai}},\
  }\bibfield  {title} {\bibinfo {title} {New classes of degree sequences with
  fast mixing swap {Markov} chain sampling},\ }\href
  {https://doi.org/10.1017/S0963548317000499} {\bibfield  {journal} {\bibinfo
  {journal} {Comb. Probab. Comput.}\ }\textbf {\bibinfo {volume} {27}},\
  \bibinfo {pages} {186} (\bibinfo {year} {2018})}\BibitemShut {NoStop}%
\bibitem [{\citenamefont {Greenhill}\ and\ \citenamefont
  {Sfragara}(2018)}]{GREENHILL20181}%
  \BibitemOpen
  \bibfield  {author} {\bibinfo {author} {\bibfnamefont {C.}~\bibnamefont
  {Greenhill}}\ and\ \bibinfo {author} {\bibfnamefont {M.}~\bibnamefont
  {Sfragara}},\ }\bibfield  {title} {\bibinfo {title} {The switch markov chain
  for sampling irregular graphs and digraphs},\ }\href
  {https://doi.org/https://doi.org/10.1016/j.tcs.2017.11.010} {\bibfield
  {journal} {\bibinfo  {journal} {Theor. Comput. Sci.}\ }\textbf {\bibinfo
  {volume} {719}},\ \bibinfo {pages} {1} (\bibinfo {year} {2018})}\BibitemShut
  {NoStop}%
\bibitem [{\citenamefont {Milo}\ \emph {et~al.}(2004)\citenamefont {Milo},
  \citenamefont {Kashtan}, \citenamefont {Itzkovitz}, \citenamefont {Newman},\
  and\ \citenamefont {Alon}}]{milo_uniform_2004}%
  \BibitemOpen
  \bibfield  {author} {\bibinfo {author} {\bibfnamefont {R.}~\bibnamefont
  {Milo}}, \bibinfo {author} {\bibfnamefont {N.}~\bibnamefont {Kashtan}},
  \bibinfo {author} {\bibfnamefont {S.}~\bibnamefont {Itzkovitz}}, \bibinfo
  {author} {\bibfnamefont {M.~E.~J.}\ \bibnamefont {Newman}},\ and\ \bibinfo
  {author} {\bibfnamefont {U.}~\bibnamefont {Alon}},\ }\href
  {https://doi.org/10.48550/arXiv.cond-mat/0312028} {\emph {\bibinfo {title}
  {On the uniform generation of random graphs with prescribed degree
  sequences}}},\ \bibinfo {type} {Preprint}\ \bibinfo {number}
  {arXiv:cond-mat/0312028}\ (\bibinfo {year} {2004})\BibitemShut {NoStop}%
\bibitem [{\citenamefont {Dutta}\ \emph {et~al.}(2023)\citenamefont {Dutta},
  \citenamefont {Fosdick},\ and\ \citenamefont
  {Clauset}}]{dutta_sampling_2023}%
  \BibitemOpen
  \bibfield  {author} {\bibinfo {author} {\bibfnamefont {U.}~\bibnamefont
  {Dutta}}, \bibinfo {author} {\bibfnamefont {B.~K.}\ \bibnamefont {Fosdick}},\
  and\ \bibinfo {author} {\bibfnamefont {A.}~\bibnamefont {Clauset}},\ }\href
  {https://doi.org/10.48550/arXiv.2105.12120} {\emph {\bibinfo {title}
  {Sampling random graphs with specified degree sequences}}},\ \bibinfo {type}
  {Preprint}\ \bibinfo {number} {arXiv:2105.12120}\ (\bibinfo {year}
  {2023})\BibitemShut {NoStop}%
\bibitem [{\citenamefont {Maslov}\ \emph {et~al.}(2004)\citenamefont {Maslov},
  \citenamefont {Sneppen},\ and\ \citenamefont {Zaliznyak}}]{MASLOV2004}%
  \BibitemOpen
  \bibfield  {author} {\bibinfo {author} {\bibfnamefont {S.}~\bibnamefont
  {Maslov}}, \bibinfo {author} {\bibfnamefont {K.}~\bibnamefont {Sneppen}},\
  and\ \bibinfo {author} {\bibfnamefont {A.}~\bibnamefont {Zaliznyak}},\
  }\bibfield  {title} {\bibinfo {title} {Detection of topological patterns in
  complex networks: correlation profile of the internet},\ }\href
  {https://doi.org/10.1016/j.physa.2003.06.002} {\bibfield  {journal} {\bibinfo
   {journal} {Physica A}\ }\textbf {\bibinfo {volume} {333}},\ \bibinfo {pages}
  {529} (\bibinfo {year} {2004})}\BibitemShut {NoStop}%
\bibitem [{\citenamefont {Stouffer}\ \emph {et~al.}(2007)\citenamefont
  {Stouffer}, \citenamefont {Camacho}, \citenamefont {Jiang},\ and\
  \citenamefont {Nunes~Amaral}}]{Stouffer_evidence_2007}%
  \BibitemOpen
  \bibfield  {author} {\bibinfo {author} {\bibfnamefont {D.~B.}\ \bibnamefont
  {Stouffer}}, \bibinfo {author} {\bibfnamefont {J.}~\bibnamefont {Camacho}},
  \bibinfo {author} {\bibfnamefont {W.}~\bibnamefont {Jiang}},\ and\ \bibinfo
  {author} {\bibfnamefont {L.~A.}\ \bibnamefont {Nunes~Amaral}},\ }\bibfield
  {title} {\bibinfo {title} {Evidence for the existence of a robust pattern of
  prey selection in food webs},\ }\href
  {https://doi.org/10.1098/rspb.2007.0571} {\bibfield  {journal} {\bibinfo
  {journal} {Proc. R. Soc. B}\ }\textbf {\bibinfo {volume} {274}},\ \bibinfo
  {pages} {1931} (\bibinfo {year} {2007})}\BibitemShut {NoStop}%
\bibitem [{\citenamefont {Chatterjee}\ \emph {et~al.}(2011)\citenamefont
  {Chatterjee}, \citenamefont {Diaconis},\ and\ \citenamefont
  {Sly}}]{Chatterjee_2011}%
  \BibitemOpen
  \bibfield  {author} {\bibinfo {author} {\bibfnamefont {S.}~\bibnamefont
  {Chatterjee}}, \bibinfo {author} {\bibfnamefont {P.}~\bibnamefont
  {Diaconis}},\ and\ \bibinfo {author} {\bibfnamefont {A.}~\bibnamefont
  {Sly}},\ }\bibfield  {title} {\bibinfo {title} {Random graphs with a given
  degree sequence},\ }\href {https://doi.org/10.1214/10-AAP728} {\bibfield
  {journal} {\bibinfo  {journal} {Ann. Appl. Probab.}\ }\textbf {\bibinfo
  {volume} {21}},\ \bibinfo {pages} {1400} (\bibinfo {year}
  {2011})}\BibitemShut {NoStop}%
\bibitem [{\citenamefont {{Miller, J. C.}}\ and\ \citenamefont {{Kiss, I.
  Z.}}(2014)}]{Miller_epidemic_2014}%
  \BibitemOpen
  \bibfield  {author} {\bibinfo {author} {\bibnamefont {{Miller, J. C.}}}\ and\
  \bibinfo {author} {\bibnamefont {{Kiss, I. Z.}}},\ }\bibfield  {title}
  {\bibinfo {title} {Epidemic spread in networks: Existing methods and current
  challenges},\ }\href {https://doi.org/10.1051/mmnp/20149202} {\bibfield
  {journal} {\bibinfo  {journal} {Math. Model. Nat. Phenom.}\ }\textbf
  {\bibinfo {volume} {9}},\ \bibinfo {pages} {4} (\bibinfo {year}
  {2014})}\BibitemShut {NoStop}%
\bibitem [{\citenamefont {{St-Onge}}\ \emph {et~al.}(2018)\citenamefont
  {{St-Onge}}, \citenamefont {Young}, \citenamefont {Laurence}, \citenamefont
  {Murphy},\ and\ \citenamefont {Dub{\'e}}}]{stonge2017susceptible}%
  \BibitemOpen
  \bibfield  {author} {\bibinfo {author} {\bibfnamefont {G.}~\bibnamefont
  {{St-Onge}}}, \bibinfo {author} {\bibfnamefont {J.-G.}\ \bibnamefont
  {Young}}, \bibinfo {author} {\bibfnamefont {E.}~\bibnamefont {Laurence}},
  \bibinfo {author} {\bibfnamefont {C.}~\bibnamefont {Murphy}},\ and\ \bibinfo
  {author} {\bibfnamefont {L.~J.}\ \bibnamefont {Dub{\'e}}},\ }\bibfield
  {title} {\bibinfo {title} {Phase transition of the
  susceptible-infected-susceptible dynamics on time-varying configuration model
  networks},\ }\href {https://doi.org/10.1103/PhysRevE.97.022305} {\bibfield
  {journal} {\bibinfo  {journal} {Phys. Rev. E}\ }\textbf {\bibinfo {volume}
  {97}},\ \bibinfo {pages} {022305} (\bibinfo {year} {2018})}\BibitemShut
  {NoStop}%
\bibitem [{\citenamefont {Orsini}\ \emph {et~al.}(2015)\citenamefont {Orsini},
  \citenamefont {Dankulov}, \citenamefont {Colomer-de Simón}, \citenamefont
  {Jamakovic}, \citenamefont {Mahadevan}, \citenamefont {Vahdat}, \citenamefont
  {Bassler}, \citenamefont {Toroczkai}, \citenamefont {Boguñá}, \citenamefont
  {Caldarelli}, \citenamefont {Fortunato},\ and\ \citenamefont
  {Krioukov}}]{orsini_quantifying_2015}%
  \BibitemOpen
  \bibfield  {author} {\bibinfo {author} {\bibfnamefont {C.}~\bibnamefont
  {Orsini}}, \bibinfo {author} {\bibfnamefont {M.~M.}\ \bibnamefont
  {Dankulov}}, \bibinfo {author} {\bibfnamefont {P.}~\bibnamefont {Colomer-de
  Simón}}, \bibinfo {author} {\bibfnamefont {A.}~\bibnamefont {Jamakovic}},
  \bibinfo {author} {\bibfnamefont {P.}~\bibnamefont {Mahadevan}}, \bibinfo
  {author} {\bibfnamefont {A.}~\bibnamefont {Vahdat}}, \bibinfo {author}
  {\bibfnamefont {K.~E.}\ \bibnamefont {Bassler}}, \bibinfo {author}
  {\bibfnamefont {Z.}~\bibnamefont {Toroczkai}}, \bibinfo {author}
  {\bibfnamefont {M.}~\bibnamefont {Boguñá}}, \bibinfo {author}
  {\bibfnamefont {G.}~\bibnamefont {Caldarelli}}, \bibinfo {author}
  {\bibfnamefont {S.}~\bibnamefont {Fortunato}},\ and\ \bibinfo {author}
  {\bibfnamefont {D.}~\bibnamefont {Krioukov}},\ }\bibfield  {title} {\bibinfo
  {title} {Quantifying randomness in real networks},\ }\href
  {https://doi.org/10.1038/ncomms9627} {\bibfield  {journal} {\bibinfo
  {journal} {Nat. Commun.}\ }\textbf {\bibinfo {volume} {6}},\ \bibinfo {pages}
  {8627} (\bibinfo {year} {2015})}\BibitemShut {NoStop}%
\bibitem [{\citenamefont {Zarei}\ \emph {et~al.}(2024)\citenamefont {Zarei},
  \citenamefont {Gandica},\ and\ \citenamefont {Rocha}}]{Zarei_bursts_2024}%
  \BibitemOpen
  \bibfield  {author} {\bibinfo {author} {\bibfnamefont {F.}~\bibnamefont
  {Zarei}}, \bibinfo {author} {\bibfnamefont {Y.}~\bibnamefont {Gandica}},\
  and\ \bibinfo {author} {\bibfnamefont {L.~E.~C.}\ \bibnamefont {Rocha}},\
  }\bibfield  {title} {\bibinfo {title} {Bursts of communication increase
  opinion diversity in the temporal deffuant model},\ }\href
  {https://doi.org/10.1038/s41598-024-52458-w} {\bibfield  {journal} {\bibinfo
  {journal} {Sci. Rep.}\ }\textbf {\bibinfo {volume} {14}},\ \bibinfo {pages}
  {2222} (\bibinfo {year} {2024})}\BibitemShut {NoStop}%
\bibitem [{\citenamefont {Okeukwu-Ogbonnaya}\ \emph {et~al.}(2024)\citenamefont
  {Okeukwu-Ogbonnaya}, \citenamefont {Amariucai}, \citenamefont {Natarajan},\
  and\ \citenamefont {Kim}}]{Okeukwu-Ogbonnaya_towards_2024}%
  \BibitemOpen
  \bibfield  {author} {\bibinfo {author} {\bibfnamefont {A.}~\bibnamefont
  {Okeukwu-Ogbonnaya}}, \bibinfo {author} {\bibfnamefont {G.}~\bibnamefont
  {Amariucai}}, \bibinfo {author} {\bibfnamefont {B.}~\bibnamefont
  {Natarajan}},\ and\ \bibinfo {author} {\bibfnamefont {H.~J.}\ \bibnamefont
  {Kim}},\ }\bibfield  {title} {\bibinfo {title} {Towards quantifying the
  communication aspect of resilience in disaster-prone communities},\ }\href
  {https://doi.org/10.1038/s41598-024-59192-3} {\bibfield  {journal} {\bibinfo
  {journal} {Sci. Rep.}\ }\textbf {\bibinfo {volume} {14}},\ \bibinfo {pages}
  {8837} (\bibinfo {year} {2024})}\BibitemShut {NoStop}%
\bibitem [{\citenamefont {Murphy}\ \emph {et~al.}(2024)\citenamefont {Murphy},
  \citenamefont {Thibeault}, \citenamefont {Allard},\ and\ \citenamefont
  {Desrosiers}}]{Murphy2024}%
  \BibitemOpen
  \bibfield  {author} {\bibinfo {author} {\bibfnamefont {C.}~\bibnamefont
  {Murphy}}, \bibinfo {author} {\bibfnamefont {V.}~\bibnamefont {Thibeault}},
  \bibinfo {author} {\bibfnamefont {A.}~\bibnamefont {Allard}},\ and\ \bibinfo
  {author} {\bibfnamefont {P.}~\bibnamefont {Desrosiers}},\ }\bibfield  {title}
  {\bibinfo {title} {Duality between predictability and reconstructability in
  complex systems},\ }\href {https://doi.org/10.1038/s41467-024-48020-x}
  {\bibfield  {journal} {\bibinfo  {journal} {Nat. Commun.}\ }\textbf {\bibinfo
  {volume} {15}},\ \bibinfo {pages} {4478} (\bibinfo {year}
  {2024})}\BibitemShut {NoStop}%
\bibitem [{\citenamefont {{Clariana-Rodagut}}\ and\ \citenamefont
  {Cardillo}(2024)}]{Clariana2024Quantifying}%
  \BibitemOpen
  \bibfield  {author} {\bibinfo {author} {\bibfnamefont {A.}~\bibnamefont
  {{Clariana-Rodagut}}}\ and\ \bibinfo {author} {\bibfnamefont
  {A.}~\bibnamefont {Cardillo}},\ }\bibfield  {title} {\bibinfo {title}
  {Quantifying {{Women}}'s {{Marginalisation}} in {{Ibero-American Film Culture
  During}} the {{First Half}} of the {{Twentieth Century}}: {{A Network-Science
  Proposal}}},\ }\href {https://doi.org/10.22148/001c.118589} {\bibfield
  {journal} {\bibinfo  {journal} {J. Cult. Anal.}\ }\textbf {\bibinfo {volume}
  {9}} (\bibinfo {year} {2024})}\BibitemShut {NoStop}%
\bibitem [{\citenamefont {Newman}(2002)}]{newman_2002_assortative}%
  \BibitemOpen
  \bibfield  {author} {\bibinfo {author} {\bibfnamefont {M.~E.~J.}\
  \bibnamefont {Newman}},\ }\bibfield  {title} {\bibinfo {title} {Assortative
  mixing in networks},\ }\href {https://doi.org/10.1103/PhysRevLett.89.208701}
  {\bibfield  {journal} {\bibinfo  {journal} {Phys. Rev. Lett.}\ }\textbf
  {\bibinfo {volume} {89}},\ \bibinfo {pages} {208701} (\bibinfo {year}
  {2002})}\BibitemShut {NoStop}%
\bibitem [{\citenamefont {Newman}(2003)}]{newman_2003_mixing}%
  \BibitemOpen
  \bibfield  {author} {\bibinfo {author} {\bibfnamefont {M.~E.~J.}\
  \bibnamefont {Newman}},\ }\bibfield  {title} {\bibinfo {title} {Mixing
  patterns in networks},\ }\href {https://doi.org/10.1103/PhysRevE.67.026126}
  {\bibfield  {journal} {\bibinfo  {journal} {Phys. Rev. E}\ }\textbf {\bibinfo
  {volume} {67}},\ \bibinfo {pages} {026126} (\bibinfo {year}
  {2003})}\BibitemShut {NoStop}%
\bibitem [{\citenamefont {Stanton}\ and\ \citenamefont
  {Pinar}(2012)}]{Stanton_2012_constructing}%
  \BibitemOpen
  \bibfield  {author} {\bibinfo {author} {\bibfnamefont {I.}~\bibnamefont
  {Stanton}}\ and\ \bibinfo {author} {\bibfnamefont {A.}~\bibnamefont
  {Pinar}},\ }\bibfield  {title} {\bibinfo {title} {Constructing and sampling
  graphs with a prescribed joint degree distribution},\ }\href
  {https://doi.org/10.1145/2133803.2330086} {\bibfield  {journal} {\bibinfo
  {journal} {ACM J. Exp. Algorithmics}\ }\textbf {\bibinfo {volume} {17}},\
  \bibinfo {pages} {3.1} (\bibinfo {year} {2012})}\BibitemShut {NoStop}%
\bibitem [{\citenamefont {Czabarka}\ \emph {et~al.}(2015)\citenamefont
  {Czabarka}, \citenamefont {Dutle}, \citenamefont {Erd{\H o}s},\ and\
  \citenamefont {Mikl{\'o}s}}]{czabarka2015realizations}%
  \BibitemOpen
  \bibfield  {author} {\bibinfo {author} {\bibfnamefont {{\'E}.}~\bibnamefont
  {Czabarka}}, \bibinfo {author} {\bibfnamefont {A.}~\bibnamefont {Dutle}},
  \bibinfo {author} {\bibfnamefont {P.~L.}\ \bibnamefont {Erd{\H o}s}},\ and\
  \bibinfo {author} {\bibfnamefont {I.}~\bibnamefont {Mikl{\'o}s}},\ }\bibfield
   {title} {\bibinfo {title} {On realizations of a joint degree matrix},\
  }\href {https://doi.org/10.1016/j.dam.2014.10.012} {\bibfield  {journal}
  {\bibinfo  {journal} {Discret. Appl. Math.}\ }\textbf {\bibinfo {volume}
  {181}},\ \bibinfo {pages} {283} (\bibinfo {year} {2015})}\BibitemShut
  {NoStop}%
\bibitem [{\citenamefont {Amanatidis}\ \emph {et~al.}(2018)\citenamefont
  {Amanatidis}, \citenamefont {Green},\ and\ \citenamefont
  {Mihail}}]{amanatidis_connected_2018}%
  \BibitemOpen
  \bibfield  {author} {\bibinfo {author} {\bibfnamefont {G.}~\bibnamefont
  {Amanatidis}}, \bibinfo {author} {\bibfnamefont {B.}~\bibnamefont {Green}},\
  and\ \bibinfo {author} {\bibfnamefont {M.}~\bibnamefont {Mihail}},\
  }\bibfield  {title} {\bibinfo {title} {Connected realizations of joint-degree
  matrices},\ }\href
  {https://doi.org/https://doi.org/10.1016/j.dam.2018.04.010} {\bibfield
  {journal} {\bibinfo  {journal} {Discret. Appl. Math.}\ }\textbf {\bibinfo
  {volume} {250}},\ \bibinfo {pages} {65} (\bibinfo {year} {2018})}\BibitemShut
  {NoStop}%
\bibitem [{\citenamefont {Erd\"{o}s}\ \emph {et~al.}(2015)\citenamefont
  {Erd\"{o}s}, \citenamefont {Miklós},\ and\ \citenamefont
  {Toroczkai}}]{erdos_2015_decomposition}%
  \BibitemOpen
  \bibfield  {author} {\bibinfo {author} {\bibfnamefont {P.~L.}\ \bibnamefont
  {Erd\"{o}s}}, \bibinfo {author} {\bibfnamefont {I.}~\bibnamefont {Miklós}},\
  and\ \bibinfo {author} {\bibfnamefont {Z.}~\bibnamefont {Toroczkai}},\
  }\bibfield  {title} {\bibinfo {title} {A decomposition based proof for fast
  mixing of a markov chain over balanced realizations of a joint degree
  matrix},\ }\href {https://doi.org/10.1137/130929874} {\bibfield  {journal}
  {\bibinfo  {journal} {SIAM J. Discrete Math.}\ }\textbf {\bibinfo {volume}
  {29}},\ \bibinfo {pages} {481} (\bibinfo {year} {2015})}\BibitemShut
  {NoStop}%
\bibitem [{\citenamefont {Stauffer}\ and\ \citenamefont
  {Barbosa}(2011)}]{stauffer_study_2011}%
  \BibitemOpen
  \bibfield  {author} {\bibinfo {author} {\bibfnamefont {A.~O.}\ \bibnamefont
  {Stauffer}}\ and\ \bibinfo {author} {\bibfnamefont {V.~C.}\ \bibnamefont
  {Barbosa}},\ }\href {https://doi.org/10.48550/arXiv.cs/0512105} {\emph
  {\bibinfo {title} {A study of the edge-switching {{Markov-chain}} method for
  the generation of random graphs}}},\ \bibinfo {type} {Preprint}\ \bibinfo
  {number} {arXiv:cs/0512105}\ (\bibinfo {year} {2011})\BibitemShut {NoStop}%
\bibitem [{\citenamefont {Ring}\ \emph {et~al.}(2020)\citenamefont {Ring},
  \citenamefont {Young},\ and\ \citenamefont
  {{H{\'e}bert-Dufresne}}}]{ring2020connected}%
  \BibitemOpen
  \bibfield  {author} {\bibinfo {author} {\bibfnamefont {J.~H.}\ \bibnamefont
  {Ring}}, \bibinfo {author} {\bibfnamefont {J.-G.}\ \bibnamefont {Young}},\
  and\ \bibinfo {author} {\bibfnamefont {L.}~\bibnamefont
  {{H{\'e}bert-Dufresne}}},\ }\bibfield  {title} {\bibinfo {title} {Connected
  {{Graphs}} with a {{Given Degree Sequence}}: {{Efficient Sampling}},
  {{Correlations}}, {{Community Detection}} and {{Robustness}}},\ }in\ \href
  {https://doi.org/10.1007/978-3-030-38965-9_3} {\emph {\bibinfo {booktitle}
  {Proceedings of {{NetSci-X}} 2020: {{Sixth International Winter School}} and
  {{Conference}} on {{Network Science}}}}}\ (\bibinfo {year} {2020})\ pp.\
  \bibinfo {pages} {33--47}\BibitemShut {NoStop}%
\bibitem [{\citenamefont {Jamakovic}\ \emph {et~al.}(2015)\citenamefont
  {Jamakovic}, \citenamefont {Mahadevan}, \citenamefont {Vahdat}, \citenamefont
  {Boguna},\ and\ \citenamefont {Krioukov}}]{jamakovic_how_2015}%
  \BibitemOpen
  \bibfield  {author} {\bibinfo {author} {\bibfnamefont {A.}~\bibnamefont
  {Jamakovic}}, \bibinfo {author} {\bibfnamefont {P.}~\bibnamefont
  {Mahadevan}}, \bibinfo {author} {\bibfnamefont {A.}~\bibnamefont {Vahdat}},
  \bibinfo {author} {\bibfnamefont {M.}~\bibnamefont {Boguna}},\ and\ \bibinfo
  {author} {\bibfnamefont {D.}~\bibnamefont {Krioukov}},\ }\href
  {https://doi.org/10.48550/arXiv.0908.1143} {\emph {\bibinfo {title} {How
  small are building blocks of complex networks}}},\ \bibinfo {type}
  {Preprint}\ \bibinfo {number} {arXiv:0908.1143}\ (\bibinfo {year}
  {2015})\BibitemShut {NoStop}%
\bibitem [{\citenamefont {Cooper}\ \emph {et~al.}(2023)\citenamefont {Cooper},
  \citenamefont {Dyer},\ and\ \citenamefont
  {Greenhill}}]{cooper_triangle_2023}%
  \BibitemOpen
  \bibfield  {author} {\bibinfo {author} {\bibfnamefont {C.}~\bibnamefont
  {Cooper}}, \bibinfo {author} {\bibfnamefont {M.}~\bibnamefont {Dyer}},\ and\
  \bibinfo {author} {\bibfnamefont {C.}~\bibnamefont {Greenhill}},\ }\href
  {http://arxiv.org/abs/2301.08499} {\emph {\bibinfo {title} {A triangle
  process on graphs with given degree sequence}}},\ \bibinfo {type} {Preprint}\
  \bibinfo {number} {arXiv:2301.08499}\ (\bibinfo {year} {2023})\BibitemShut
  {NoStop}%
\bibitem [{\citenamefont {Stamm}\ \emph {et~al.}(2023)\citenamefont {Stamm},
  \citenamefont {Scholkemper}, \citenamefont {Strohmaier},\ and\ \citenamefont
  {Schaub}}]{Stamm_2023_neighborhood}%
  \BibitemOpen
  \bibfield  {author} {\bibinfo {author} {\bibfnamefont {F.~I.}\ \bibnamefont
  {Stamm}}, \bibinfo {author} {\bibfnamefont {M.}~\bibnamefont {Scholkemper}},
  \bibinfo {author} {\bibfnamefont {M.}~\bibnamefont {Strohmaier}},\ and\
  \bibinfo {author} {\bibfnamefont {M.~T.}\ \bibnamefont {Schaub}},\ }\bibfield
   {title} {\bibinfo {title} {Neighborhood structure configuration models},\
  }in\ \href {https://doi.org/10.1145/3543507.3583266} {\emph {\bibinfo
  {booktitle} {Proceedings of the ACM Web Conference}}}\ (\bibinfo {year}
  {2023})\ p.\ \bibinfo {pages} {210–220}\BibitemShut {NoStop}%
\bibitem [{\citenamefont {Van~Koevering}\ \emph {et~al.}(2021)\citenamefont
  {Van~Koevering}, \citenamefont {Benson},\ and\ \citenamefont
  {Kleinberg}}]{van_koevering_random_2021}%
  \BibitemOpen
  \bibfield  {author} {\bibinfo {author} {\bibfnamefont {K.}~\bibnamefont
  {Van~Koevering}}, \bibinfo {author} {\bibfnamefont {A.}~\bibnamefont
  {Benson}},\ and\ \bibinfo {author} {\bibfnamefont {J.}~\bibnamefont
  {Kleinberg}},\ }\bibfield  {title} {\bibinfo {title} {Random {Graphs} with
  {Prescribed} {K}-{Core} {Sequences}: {A} {New} {Null} {Model} for {Network}
  {Analysis}},\ }in\ \href {https://doi.org/10.1145/3442381.3450001} {\emph
  {\bibinfo {booktitle} {Proceedings of the {Web} {Conference}}}}\ (\bibinfo
  {year} {2021})\ pp.\ \bibinfo {pages} {367--378}\BibitemShut {NoStop}%
\bibitem [{\citenamefont {Hébert-Dufresne}\ \emph {et~al.}(2016)\citenamefont
  {Hébert-Dufresne}, \citenamefont {Grochow},\ and\ \citenamefont
  {Allard}}]{hebert-dufresne_multi-scale_2016}%
  \BibitemOpen
  \bibfield  {author} {\bibinfo {author} {\bibfnamefont {L.}~\bibnamefont
  {Hébert-Dufresne}}, \bibinfo {author} {\bibfnamefont {J.~A.}\ \bibnamefont
  {Grochow}},\ and\ \bibinfo {author} {\bibfnamefont {A.}~\bibnamefont
  {Allard}},\ }\bibfield  {title} {\bibinfo {title} {Multi-scale structure and
  topological anomaly detection via a new network statistic: {The} onion
  decomposition},\ }\href {https://doi.org/10.1038/srep31708} {\bibfield
  {journal} {\bibinfo  {journal} {Sci. Rep.}\ }\textbf {\bibinfo {volume}
  {6}},\ \bibinfo {pages} {31708} (\bibinfo {year} {2016})}\BibitemShut
  {NoStop}%
\bibitem [{\citenamefont {Kong}\ \emph {et~al.}(2019)\citenamefont {Kong},
  \citenamefont {Shi}, \citenamefont {Wu},\ and\ \citenamefont
  {Zhang}}]{kong_kmathmi_2019}%
  \BibitemOpen
  \bibfield  {author} {\bibinfo {author} {\bibfnamefont {Y.-X.}\ \bibnamefont
  {Kong}}, \bibinfo {author} {\bibfnamefont {G.-Y.}\ \bibnamefont {Shi}},
  \bibinfo {author} {\bibfnamefont {R.-J.}\ \bibnamefont {Wu}},\ and\ \bibinfo
  {author} {\bibfnamefont {Y.-C.}\ \bibnamefont {Zhang}},\ }\bibfield  {title}
  {\bibinfo {title} {k-core: {{Theories}} and applications},\ }\href
  {https://doi.org/10.1016/j.physrep.2019.10.004} {\bibfield  {journal}
  {\bibinfo  {journal} {Phys. Rep.}\ }\textbf {\bibinfo {volume} {832}},\
  \bibinfo {pages} {1} (\bibinfo {year} {2019})}\BibitemShut {NoStop}%
\bibitem [{\citenamefont {Kitsak}\ \emph {et~al.}(2010)\citenamefont {Kitsak},
  \citenamefont {Gallos}, \citenamefont {Havlin}, \citenamefont {Liljeros},
  \citenamefont {Muchnik}, \citenamefont {Stanley},\ and\ \citenamefont
  {Makse}}]{kitsak2010identification}%
  \BibitemOpen
  \bibfield  {author} {\bibinfo {author} {\bibfnamefont {M.}~\bibnamefont
  {Kitsak}}, \bibinfo {author} {\bibfnamefont {L.~K.}\ \bibnamefont {Gallos}},
  \bibinfo {author} {\bibfnamefont {S.}~\bibnamefont {Havlin}}, \bibinfo
  {author} {\bibfnamefont {F.}~\bibnamefont {Liljeros}}, \bibinfo {author}
  {\bibfnamefont {L.}~\bibnamefont {Muchnik}}, \bibinfo {author} {\bibfnamefont
  {H.~E.}\ \bibnamefont {Stanley}},\ and\ \bibinfo {author} {\bibfnamefont
  {H.~A.}\ \bibnamefont {Makse}},\ }\bibfield  {title} {\bibinfo {title}
  {Identification of influential spreaders in complex networks},\ }\href
  {https://doi.org/10.1038/nphys1746} {\bibfield  {journal} {\bibinfo
  {journal} {Nat. Phys.}\ }\textbf {\bibinfo {volume} {6}},\ \bibinfo {pages}
  {888} (\bibinfo {year} {2010})}\BibitemShut {NoStop}%
\bibitem [{\citenamefont {Batagelj}\ and\ \citenamefont {Zaver{\v
  s}nik}(2011)}]{batagelj2011fast}%
  \BibitemOpen
  \bibfield  {author} {\bibinfo {author} {\bibfnamefont {V.}~\bibnamefont
  {Batagelj}}\ and\ \bibinfo {author} {\bibfnamefont {M.}~\bibnamefont
  {Zaver{\v s}nik}},\ }\bibfield  {title} {\bibinfo {title} {Fast algorithms
  for determining (generalized) core groups in social networks},\ }\href
  {https://doi.org/10.1007/s11634-010-0079-y} {\bibfield  {journal} {\bibinfo
  {journal} {Adv. Data Anal. Classif.}\ }\textbf {\bibinfo {volume} {5}},\
  \bibinfo {pages} {129} (\bibinfo {year} {2011})}\BibitemShut {NoStop}%
\bibitem [{\citenamefont {Young}\ \emph {et~al.}(2019)\citenamefont {Young},
  \citenamefont {St-Onge}, \citenamefont {Laurence}, \citenamefont {Murphy},
  \citenamefont {H{\'e}bert-Dufresne},\ and\ \citenamefont
  {Desrosiers}}]{young2019phase}%
  \BibitemOpen
  \bibfield  {author} {\bibinfo {author} {\bibfnamefont {J.-G.}\ \bibnamefont
  {Young}}, \bibinfo {author} {\bibfnamefont {G.}~\bibnamefont {St-Onge}},
  \bibinfo {author} {\bibfnamefont {E.}~\bibnamefont {Laurence}}, \bibinfo
  {author} {\bibfnamefont {C.}~\bibnamefont {Murphy}}, \bibinfo {author}
  {\bibfnamefont {L.}~\bibnamefont {H{\'e}bert-Dufresne}},\ and\ \bibinfo
  {author} {\bibfnamefont {P.}~\bibnamefont {Desrosiers}},\ }\bibfield  {title}
  {\bibinfo {title} {Phase transition in the recoverability of network
  history},\ }\href {https://doi.org/10.1103/PhysRevX.9.041056} {\bibfield
  {journal} {\bibinfo  {journal} {Phys. Rev. X}\ }\textbf {\bibinfo {volume}
  {9}},\ \bibinfo {pages} {041056} (\bibinfo {year} {2019})}\BibitemShut
  {NoStop}%
\bibitem [{\citenamefont {Mimar}\ and\ \citenamefont
  {Ghoshal}(2022)}]{mimar2022sampling}%
  \BibitemOpen
  \bibfield  {author} {\bibinfo {author} {\bibfnamefont {S.}~\bibnamefont
  {Mimar}}\ and\ \bibinfo {author} {\bibfnamefont {G.}~\bibnamefont
  {Ghoshal}},\ }\bibfield  {title} {\bibinfo {title} {A sampling-guided
  unsupervised learning method to capture percolation in complex networks},\
  }\href {https://doi.org/10.1038/s41598-022-07921-x} {\bibfield  {journal}
  {\bibinfo  {journal} {Sci. Rep.}\ }\textbf {\bibinfo {volume} {12}},\
  \bibinfo {pages} {4147} (\bibinfo {year} {2022})}\BibitemShut {NoStop}%
\bibitem [{\citenamefont {Lu}\ \emph {et~al.}(2024)\citenamefont {Lu},
  \citenamefont {Huang}, \citenamefont {Nie}, \citenamefont {Chen},\ and\
  \citenamefont {Xuan}}]{lu2024rk}%
  \BibitemOpen
  \bibfield  {author} {\bibinfo {author} {\bibfnamefont {Y.}~\bibnamefont
  {Lu}}, \bibinfo {author} {\bibfnamefont {Y.}~\bibnamefont {Huang}}, \bibinfo
  {author} {\bibfnamefont {J.}~\bibnamefont {Nie}}, \bibinfo {author}
  {\bibfnamefont {Z.}~\bibnamefont {Chen}},\ and\ \bibinfo {author}
  {\bibfnamefont {Q.}~\bibnamefont {Xuan}},\ }\bibfield  {title} {\bibinfo
  {title} {{{RK-CORE}}: {{An Established Methodology}} for {{Exploring}} the
  {{Hierarchical Structure}} within {{Datasets}}},\ }in\ \href
  {https://doi.org/10.1109/ICASSP48485.2024.10447791} {\emph {\bibinfo
  {booktitle} {{{IEEE International Conference}} on {{Acoustics}}, {{Speech}}
  and {{Signal Processing}} ({{ICASSP}})}}}\ (\bibinfo {year} {2024})\ pp.\
  \bibinfo {pages} {3150--3154}\BibitemShut {NoStop}%
\bibitem [{\citenamefont {{Garc{\'i}a-P{\'e}rez}}\ \emph
  {et~al.}(2019)\citenamefont {{Garc{\'i}a-P{\'e}rez}}, \citenamefont {Allard},
  \citenamefont {Serrano},\ and\ \citenamefont
  {Bogu{\~n}{\'a}}}]{garcia-perez2019mercator}%
  \BibitemOpen
  \bibfield  {author} {\bibinfo {author} {\bibfnamefont {G.}~\bibnamefont
  {{Garc{\'i}a-P{\'e}rez}}}, \bibinfo {author} {\bibfnamefont {A.}~\bibnamefont
  {Allard}}, \bibinfo {author} {\bibfnamefont {M.~{\'A}.}\ \bibnamefont
  {Serrano}},\ and\ \bibinfo {author} {\bibfnamefont {M.}~\bibnamefont
  {Bogu{\~n}{\'a}}},\ }\bibfield  {title} {\bibinfo {title} {Mercator:
  uncovering faithful hyperbolic embeddings of complex networks},\ }\href
  {https://doi.org/10.1088/1367-2630/ab57d2} {\bibfield  {journal} {\bibinfo
  {journal} {New J. Phys.}\ }\textbf {\bibinfo {volume} {21}},\ \bibinfo
  {pages} {123033} (\bibinfo {year} {2019})}\BibitemShut {NoStop}%
\bibitem [{\citenamefont {Zhou}\ \emph {et~al.}(2023)\citenamefont {Zhou},
  \citenamefont {Duenas-Osorio}, \citenamefont {Doss-Gollin}, \citenamefont
  {Liu}, \citenamefont {Stadler},\ and\ \citenamefont
  {Li}}]{zhou2023mesoscale}%
  \BibitemOpen
  \bibfield  {author} {\bibinfo {author} {\bibfnamefont {X.}~\bibnamefont
  {Zhou}}, \bibinfo {author} {\bibfnamefont {L.}~\bibnamefont {Duenas-Osorio}},
  \bibinfo {author} {\bibfnamefont {J.}~\bibnamefont {Doss-Gollin}}, \bibinfo
  {author} {\bibfnamefont {L.}~\bibnamefont {Liu}}, \bibinfo {author}
  {\bibfnamefont {L.}~\bibnamefont {Stadler}},\ and\ \bibinfo {author}
  {\bibfnamefont {Q.}~\bibnamefont {Li}},\ }\bibfield  {title} {\bibinfo
  {title} {Mesoscale modeling of distributed water systems enables policy
  search},\ }\href {https://doi.org/10.1029/2022WR033758} {\bibfield  {journal}
  {\bibinfo  {journal} {Water Resour. Res.}\ }\textbf {\bibinfo {volume}
  {59}},\ \bibinfo {pages} {e2022WR033758} (\bibinfo {year}
  {2023})}\BibitemShut {NoStop}%
\bibitem [{\citenamefont {H{\'e}bert-Dufresne}\ \emph
  {et~al.}(2023)\citenamefont {H{\'e}bert-Dufresne}, \citenamefont {St-Onge},
  \citenamefont {Meluso}, \citenamefont {Bagrow},\ and\ \citenamefont
  {Allard}}]{hebert2023hierarchical}%
  \BibitemOpen
  \bibfield  {author} {\bibinfo {author} {\bibfnamefont {L.}~\bibnamefont
  {H{\'e}bert-Dufresne}}, \bibinfo {author} {\bibfnamefont {G.}~\bibnamefont
  {St-Onge}}, \bibinfo {author} {\bibfnamefont {J.}~\bibnamefont {Meluso}},
  \bibinfo {author} {\bibfnamefont {J.}~\bibnamefont {Bagrow}},\ and\ \bibinfo
  {author} {\bibfnamefont {A.}~\bibnamefont {Allard}},\ }\bibfield  {title}
  {\bibinfo {title} {Hierarchical team structure and multidimensional
  localization (or siloing) on networks},\ }\href
  {https://doi.org/10.1088/2632-072X/ace602} {\bibfield  {journal} {\bibinfo
  {journal} {J. Phys. Complex.}\ }\textbf {\bibinfo {volume} {4}},\ \bibinfo
  {pages} {035002} (\bibinfo {year} {2023})}\BibitemShut {NoStop}%
\bibitem [{\citenamefont {Bonnaire}\ \emph {et~al.}(2020)\citenamefont
  {Bonnaire}, \citenamefont {Aghanim}, \citenamefont {Decelle},\ and\
  \citenamefont {Douspis}}]{bonnaire2020trex}%
  \BibitemOpen
  \bibfield  {author} {\bibinfo {author} {\bibfnamefont {T.}~\bibnamefont
  {Bonnaire}}, \bibinfo {author} {\bibfnamefont {N.}~\bibnamefont {Aghanim}},
  \bibinfo {author} {\bibfnamefont {A.}~\bibnamefont {Decelle}},\ and\ \bibinfo
  {author} {\bibfnamefont {M.}~\bibnamefont {Douspis}},\ }\bibfield  {title}
  {\bibinfo {title} {T-{{ReX}}: a graph-based filament detection method},\
  }\href {https://doi.org/10.1051/0004-6361/201936859} {\bibfield  {journal}
  {\bibinfo  {journal} {Astron. Astrophys.}\ }\textbf {\bibinfo {volume}
  {637}},\ \bibinfo {pages} {A18} (\bibinfo {year} {2020})}\BibitemShut
  {NoStop}%
\bibitem [{\citenamefont {Allard}\ and\ \citenamefont
  {Hébert-Dufresne}(2019)}]{allard_percolation_2019}%
  \BibitemOpen
  \bibfield  {author} {\bibinfo {author} {\bibfnamefont {A.}~\bibnamefont
  {Allard}}\ and\ \bibinfo {author} {\bibfnamefont {L.}~\bibnamefont
  {Hébert-Dufresne}},\ }\bibfield  {title} {\bibinfo {title} {Percolation and
  the {Effective} {Structure} of {Complex} {Networks}},\ }\href
  {https://doi.org/10.1103/PhysRevX.9.011023} {\bibfield  {journal} {\bibinfo
  {journal} {Phys. Rev. X}\ }\textbf {\bibinfo {volume} {9}},\ \bibinfo {pages}
  {011023} (\bibinfo {year} {2019})}\BibitemShut {NoStop}%
\bibitem [{\citenamefont {H\'ebert-Dufresne}\ \emph {et~al.}(2024)\citenamefont
  {H\'ebert-Dufresne}, \citenamefont {Young}, \citenamefont {Daniels},
  \citenamefont {Kirkley},\ and\ \citenamefont
  {Allard}}]{hebertdufresne_2024_network}%
  \BibitemOpen
  \bibfield  {author} {\bibinfo {author} {\bibfnamefont {L.}~\bibnamefont
  {H\'ebert-Dufresne}}, \bibinfo {author} {\bibfnamefont {J.-G.}\ \bibnamefont
  {Young}}, \bibinfo {author} {\bibfnamefont {A.}~\bibnamefont {Daniels}},
  \bibinfo {author} {\bibfnamefont {A.}~\bibnamefont {Kirkley}},\ and\ \bibinfo
  {author} {\bibfnamefont {A.}~\bibnamefont {Allard}},\ }\bibfield  {title}
  {\bibinfo {title} {Network compression with configuration models and the
  minimum description length},\ }\href
  {https://doi.org/10.1103/PhysRevE.110.034305} {\bibfield  {journal} {\bibinfo
   {journal} {Phys. Rev. E}\ }\textbf {\bibinfo {volume} {110}},\ \bibinfo
  {pages} {034305} (\bibinfo {year} {2024})}\BibitemShut {NoStop}%
\bibitem [{\citenamefont {Miklós}\ and\ \citenamefont
  {Podani}(2004)}]{miklos_randomization_2004}%
  \BibitemOpen
  \bibfield  {author} {\bibinfo {author} {\bibfnamefont {I.}~\bibnamefont
  {Miklós}}\ and\ \bibinfo {author} {\bibfnamefont {J.}~\bibnamefont
  {Podani}},\ }\bibfield  {title} {\bibinfo {title} {Randomization of
  {Presence}–{Absence} {Matrices}: {Comments} and {New} {Algorithms}},\
  }\href {https://doi.org/10.1890/03-0101} {\bibfield  {journal} {\bibinfo
  {journal} {Ecology}\ }\textbf {\bibinfo {volume} {85}},\ \bibinfo {pages}
  {86} (\bibinfo {year} {2004})}\BibitemShut {NoStop}%
\bibitem [{\citenamefont {Artzy-Randrup}\ and\ \citenamefont
  {Stone}(2005)}]{artzyrandrup_2005_generating}%
  \BibitemOpen
  \bibfield  {author} {\bibinfo {author} {\bibfnamefont {Y.}~\bibnamefont
  {Artzy-Randrup}}\ and\ \bibinfo {author} {\bibfnamefont {L.}~\bibnamefont
  {Stone}},\ }\bibfield  {title} {\bibinfo {title} {Generating uniformly
  distributed random networks},\ }\href
  {https://doi.org/10.1103/PhysRevE.72.056708} {\bibfield  {journal} {\bibinfo
  {journal} {Phys. Rev. E}\ }\textbf {\bibinfo {volume} {72}},\ \bibinfo
  {pages} {056708} (\bibinfo {year} {2005})}\BibitemShut {NoStop}%
\bibitem [{Note1()}]{Note1}%
  \BibitemOpen
  \bibinfo {note} {This will always happen for any graph with a node $u$ of
  degree 2 or more, since we can always take its two edges $(u,v)$ and $(u,x)$
  and swap them. This will either end up with the same graph, or be refused
  because it creates a self-loop. In both cases, this is a self-loop in the
  configuration graph. If there is no node of degree $2$ or more, then all
  nodes are of degree $1$ and of layer $1$. In this case, the configuration
  graph of the LCM is equivalent to the one from the CM, which we know to be
  aperiodic~\cite {fosdick_configuring_2018}.}\BibitemShut {Stop}%
\bibitem [{\citenamefont {Preti}\ \emph {et~al.}(2024)\citenamefont {Preti},
  \citenamefont {De~Francisci~Morales},\ and\ \citenamefont
  {Riondato}}]{preti_impossibility_2024}%
  \BibitemOpen
  \bibfield  {author} {\bibinfo {author} {\bibfnamefont {G.}~\bibnamefont
  {Preti}}, \bibinfo {author} {\bibfnamefont {G.}~\bibnamefont
  {De~Francisci~Morales}},\ and\ \bibinfo {author} {\bibfnamefont
  {M.}~\bibnamefont {Riondato}},\ }\bibfield  {title} {\bibinfo {title}
  {Impossibility result for {{Markov}} chain {{Monte Carlo}} sampling from
  microcanonical bipartite graph ensembles},\ }\href
  {https://doi.org/10.1103/PhysRevE.109.L053301} {\bibfield  {journal}
  {\bibinfo  {journal} {Phys. Rev. E}\ }\textbf {\bibinfo {volume} {109}},\
  \bibinfo {pages} {L053301} (\bibinfo {year} {2024})}\BibitemShut {NoStop}%
\bibitem [{\citenamefont {Gkantsidis}\ \emph {et~al.}(2003)\citenamefont
  {Gkantsidis}, \citenamefont {Mihail},\ and\ \citenamefont
  {Zegura}}]{gkantsidis_markov_2003}%
  \BibitemOpen
  \bibfield  {author} {\bibinfo {author} {\bibfnamefont {C.}~\bibnamefont
  {Gkantsidis}}, \bibinfo {author} {\bibfnamefont {M.}~\bibnamefont {Mihail}},\
  and\ \bibinfo {author} {\bibfnamefont {E.}~\bibnamefont {Zegura}},\
  }\bibfield  {title} {\bibinfo {title} {The {Markov} {Chain} {Simulation}
  {Method} for {Generating} {Connected} {Power} {Law} {Random} {Graphs}},\ }in\
  \href {https://dblp.org/db/conf/alenex/alenex2003.html} {\emph {\bibinfo
  {booktitle} {{Proceedings of the Fifth Workshop on Algorithm Engineering and
  Experiments}}}}\ (\bibinfo {year} {2003})\ pp.\ \bibinfo {pages}
  {16--25}\BibitemShut {NoStop}%
\bibitem [{\citenamefont {Viger}\ and\ \citenamefont
  {Latapy}(2005)}]{viger_efficient_2005}%
  \BibitemOpen
  \bibfield  {author} {\bibinfo {author} {\bibfnamefont {F.}~\bibnamefont
  {Viger}}\ and\ \bibinfo {author} {\bibfnamefont {M.}~\bibnamefont {Latapy}},\
  }\bibfield  {title} {\bibinfo {title} {Efficient and {Simple} {Generation} of
  {Random} {Simple} {Connected} {Graphs} with {Prescribed} {Degree}
  {Sequence}},\ }in\ \href {https://doi.org/10.1007/11533719_45} {\emph
  {\bibinfo {booktitle} {Computing and {Combinatorics}}}}\ (\bibinfo {year}
  {2005})\ pp.\ \bibinfo {pages} {440--449}\BibitemShut {NoStop}%
\bibitem [{\citenamefont {Tabourier}\ \emph {et~al.}(2011)\citenamefont
  {Tabourier}, \citenamefont {Roth},\ and\ \citenamefont
  {Cointet}}]{tabourier_generating_2011}%
  \BibitemOpen
  \bibfield  {author} {\bibinfo {author} {\bibfnamefont {L.}~\bibnamefont
  {Tabourier}}, \bibinfo {author} {\bibfnamefont {C.}~\bibnamefont {Roth}},\
  and\ \bibinfo {author} {\bibfnamefont {J.-P.}\ \bibnamefont {Cointet}},\
  }\bibfield  {title} {\bibinfo {title} {Generating constrained random graphs
  using multiple edge switches},\ }\href
  {https://doi.org/10.1145/1963190.2063515} {\bibfield  {journal} {\bibinfo
  {journal} {ACM J. Exp. Algorithmics}\ }\textbf {\bibinfo {volume} {16}},\
  \bibinfo {pages} {1.7:1.1} (\bibinfo {year} {2011})}\BibitemShut {NoStop}%
\bibitem [{\citenamefont {Martinez}(1991)}]{Martinez_littlerock_1991}%
  \BibitemOpen
  \bibfield  {author} {\bibinfo {author} {\bibfnamefont {N.~D.}\ \bibnamefont
  {Martinez}},\ }\bibfield  {title} {\bibinfo {title} {Artifacts or attributes?
  effects of resolution on the little rock lake food web},\ }\href
  {https://doi.org/https://doi.org/10.2307/2937047} {\bibfield  {journal}
  {\bibinfo  {journal} {Ecol. Monogr.}\ }\textbf {\bibinfo {volume} {61}},\
  \bibinfo {pages} {367} (\bibinfo {year} {1991})}\BibitemShut {NoStop}%
\bibitem [{\citenamefont {Cook}\ \emph {et~al.}(2019)\citenamefont {Cook},
  \citenamefont {Jarrell}, \citenamefont {Brittin}, \citenamefont {Wang},
  \citenamefont {Bloniarz}, \citenamefont {Yakovlev}, \citenamefont {Nguyen},
  \citenamefont {Tang}, \citenamefont {Bayer}, \citenamefont {Duerr},
  \citenamefont {B{\"u}low}, \citenamefont {Hobert}, \citenamefont {Hall},\
  and\ \citenamefont {Emmons}}]{Cook2019_male_chemical_corrected}%
  \BibitemOpen
  \bibfield  {author} {\bibinfo {author} {\bibfnamefont {S.~J.}\ \bibnamefont
  {Cook}}, \bibinfo {author} {\bibfnamefont {T.~A.}\ \bibnamefont {Jarrell}},
  \bibinfo {author} {\bibfnamefont {C.~A.}\ \bibnamefont {Brittin}}, \bibinfo
  {author} {\bibfnamefont {Y.}~\bibnamefont {Wang}}, \bibinfo {author}
  {\bibfnamefont {A.~E.}\ \bibnamefont {Bloniarz}}, \bibinfo {author}
  {\bibfnamefont {M.~A.}\ \bibnamefont {Yakovlev}}, \bibinfo {author}
  {\bibfnamefont {K.~C.~Q.}\ \bibnamefont {Nguyen}}, \bibinfo {author}
  {\bibfnamefont {L.~T.-H.}\ \bibnamefont {Tang}}, \bibinfo {author}
  {\bibfnamefont {E.~A.}\ \bibnamefont {Bayer}}, \bibinfo {author}
  {\bibfnamefont {J.~S.}\ \bibnamefont {Duerr}}, \bibinfo {author}
  {\bibfnamefont {H.~E.}\ \bibnamefont {B{\"u}low}}, \bibinfo {author}
  {\bibfnamefont {O.}~\bibnamefont {Hobert}}, \bibinfo {author} {\bibfnamefont
  {D.~H.}\ \bibnamefont {Hall}},\ and\ \bibinfo {author} {\bibfnamefont
  {S.~W.}\ \bibnamefont {Emmons}},\ }\bibfield  {title} {\bibinfo {title}
  {Whole-animal connectomes of both caenorhabditis elegans sexes},\ }\href
  {https://doi.org/10.1038/s41586-019-1352-7} {\bibfield  {journal} {\bibinfo
  {journal} {Nature}\ }\textbf {\bibinfo {volume} {571}},\ \bibinfo {pages}
  {63} (\bibinfo {year} {2019})}\BibitemShut {NoStop}%
\bibitem [{\citenamefont {Thibeault}\ \emph {et~al.}(2024)\citenamefont
  {Thibeault}, \citenamefont {Allard},\ and\ \citenamefont
  {Desrosiers}}]{Thibeault2024}%
  \BibitemOpen
  \bibfield  {author} {\bibinfo {author} {\bibfnamefont {V.}~\bibnamefont
  {Thibeault}}, \bibinfo {author} {\bibfnamefont {A.}~\bibnamefont {Allard}},\
  and\ \bibinfo {author} {\bibfnamefont {P.}~\bibnamefont {Desrosiers}},\
  }\bibfield  {title} {\bibinfo {title} {The low-rank hypothesis of complex
  systems},\ }\href {https://doi.org/10.1038/s41567-023-02303-0} {\bibfield
  {journal} {\bibinfo  {journal} {Nat. Phys.}\ }\textbf {\bibinfo {volume}
  {20}},\ \bibinfo {pages} {294} (\bibinfo {year} {2024})}\BibitemShut
  {NoStop}%
\bibitem [{\citenamefont {Miller}\ and\ \citenamefont
  {Ting}(2019)}]{Miller2019}%
  \BibitemOpen
  \bibfield  {author} {\bibinfo {author} {\bibfnamefont {J.~C.}\ \bibnamefont
  {Miller}}\ and\ \bibinfo {author} {\bibfnamefont {T.}~\bibnamefont {Ting}},\
  }\bibfield  {title} {\bibinfo {title} {{{EoN}} ({{Epidemics}} on
  {{Networks}}): a fast, flexible {{Python}} package for simulation, analytic
  approximation, and analysis of epidemics on networks},\ }\href
  {https://doi.org/10.21105/joss.01731} {\bibfield  {journal} {\bibinfo
  {journal} {J. Open Source Softw.}\ }\textbf {\bibinfo {volume} {4}},\
  \bibinfo {pages} {1731} (\bibinfo {year} {2019})}\BibitemShut {NoStop}%
\bibitem [{\citenamefont {{St-Onge}}\ \emph {et~al.}(2019)\citenamefont
  {{St-Onge}}, \citenamefont {Young}, \citenamefont {{H{\'e}bert-Dufresne}},\
  and\ \citenamefont {Dub{\'e}}}]{st-onge2019efficient}%
  \BibitemOpen
  \bibfield  {author} {\bibinfo {author} {\bibfnamefont {G.}~\bibnamefont
  {{St-Onge}}}, \bibinfo {author} {\bibfnamefont {J.-G.}\ \bibnamefont
  {Young}}, \bibinfo {author} {\bibfnamefont {L.}~\bibnamefont
  {{H{\'e}bert-Dufresne}}},\ and\ \bibinfo {author} {\bibfnamefont {L.~J.}\
  \bibnamefont {Dub{\'e}}},\ }\bibfield  {title} {\bibinfo {title} {Efficient
  sampling of spreading processes on complex networks using a composition and
  rejection algorithm},\ }\href {https://doi.org/10.1016/J.CPC.2019.02.008}
  {\bibfield  {journal} {\bibinfo  {journal} {Comput. Phys. Commun.}\ }\textbf
  {\bibinfo {volume} {240}},\ \bibinfo {pages} {30} (\bibinfo {year}
  {2019})}\BibitemShut {NoStop}%
\bibitem [{\citenamefont {Fire}\ and\ \citenamefont
  {Puzis}(2016)}]{Fire2016_facebookL1}%
  \BibitemOpen
  \bibfield  {author} {\bibinfo {author} {\bibfnamefont {M.}~\bibnamefont
  {Fire}}\ and\ \bibinfo {author} {\bibfnamefont {R.}~\bibnamefont {Puzis}},\
  }\bibfield  {title} {\bibinfo {title} {Organization mining using online
  social networks},\ }\href {https://doi.org/10.1007/s11067-015-9288-4}
  {\bibfield  {journal} {\bibinfo  {journal} {Netw. Spat. Econ.}\ }\textbf
  {\bibinfo {volume} {16}},\ \bibinfo {pages} {545} (\bibinfo {year}
  {2016})}\BibitemShut {NoStop}%
\bibitem [{\citenamefont {Watts}\ and\ \citenamefont
  {Strogatz}(1998)}]{Watts1998_celegansneural}%
  \BibitemOpen
  \bibfield  {author} {\bibinfo {author} {\bibfnamefont {D.~J.}\ \bibnamefont
  {Watts}}\ and\ \bibinfo {author} {\bibfnamefont {S.~H.}\ \bibnamefont
  {Strogatz}},\ }\bibfield  {title} {\bibinfo {title} {Collective dynamics of
  `small-world' networks},\ }\href {https://doi.org/10.1038/30918} {\bibfield
  {journal} {\bibinfo  {journal} {Nature}\ }\textbf {\bibinfo {volume} {393}},\
  \bibinfo {pages} {440} (\bibinfo {year} {1998})}\BibitemShut {NoStop}%
\bibitem [{\citenamefont {Gupta}\ \emph {et~al.}(1989)\citenamefont {Gupta},
  \citenamefont {Anderson},\ and\ \citenamefont {May}}]{Gupta1989NetworksOS}%
  \BibitemOpen
  \bibfield  {author} {\bibinfo {author} {\bibfnamefont {S.}~\bibnamefont
  {Gupta}}, \bibinfo {author} {\bibfnamefont {R.~M.}\ \bibnamefont
  {Anderson}},\ and\ \bibinfo {author} {\bibfnamefont {R.~M.}\ \bibnamefont
  {May}},\ }\bibfield  {title} {\bibinfo {title} {Networks of sexual contacts:
  implications for the pattern of spread of hiv},\ }\href
  {https://api.semanticscholar.org/CorpusID:7000357} {\bibfield  {journal}
  {\bibinfo  {journal} {AIDS}\ }\textbf {\bibinfo {volume} {3}},\ \bibinfo
  {pages} {807–818} (\bibinfo {year} {1989})}\BibitemShut {NoStop}%
\bibitem [{\citenamefont {Van~Mieghem}\ \emph {et~al.}(2010)\citenamefont
  {Van~Mieghem}, \citenamefont {Wang}, \citenamefont {Ge}, \citenamefont
  {Tang},\ and\ \citenamefont {Kuipers}}]{VanMieghem2010}%
  \BibitemOpen
  \bibfield  {author} {\bibinfo {author} {\bibfnamefont {P.}~\bibnamefont
  {Van~Mieghem}}, \bibinfo {author} {\bibfnamefont {H.}~\bibnamefont {Wang}},
  \bibinfo {author} {\bibfnamefont {X.}~\bibnamefont {Ge}}, \bibinfo {author}
  {\bibfnamefont {S.}~\bibnamefont {Tang}},\ and\ \bibinfo {author}
  {\bibfnamefont {F.~A.}\ \bibnamefont {Kuipers}},\ }\bibfield  {title}
  {\bibinfo {title} {Influence of assortativity and degree-preserving rewiring
  on the spectra of networks},\ }\href
  {https://doi.org/10.1140/epjb/e2010-00219-x} {\bibfield  {journal} {\bibinfo
  {journal} {Eur. Phys. J. B}\ }\textbf {\bibinfo {volume} {76}},\ \bibinfo
  {pages} {643} (\bibinfo {year} {2010})}\BibitemShut {NoStop}%
\bibitem [{\citenamefont {Mata}\ \emph {et~al.}(2014)\citenamefont {Mata},
  \citenamefont {Ferreira},\ and\ \citenamefont {Ferreira}}]{Mata_2014}%
  \BibitemOpen
  \bibfield  {author} {\bibinfo {author} {\bibfnamefont {A.~S.}\ \bibnamefont
  {Mata}}, \bibinfo {author} {\bibfnamefont {R.~S.}\ \bibnamefont {Ferreira}},\
  and\ \bibinfo {author} {\bibfnamefont {S.~C.}\ \bibnamefont {Ferreira}},\
  }\bibfield  {title} {\bibinfo {title} {Heterogeneous pair-approximation for
  the contact process on complex networks},\ }\href
  {https://doi.org/10.1088/1367-2630/16/5/053006} {\bibfield  {journal}
  {\bibinfo  {journal} {New J. Phys.}\ }\textbf {\bibinfo {volume} {16}},\
  \bibinfo {pages} {053006} (\bibinfo {year} {2014})}\BibitemShut {NoStop}%
\bibitem [{\citenamefont {White}\ \emph {et~al.}(1986)\citenamefont {White},
  \citenamefont {Southgate}, \citenamefont {Thomson},\ and\ \citenamefont
  {Brenner}}]{White_celegansneural}%
  \BibitemOpen
  \bibfield  {author} {\bibinfo {author} {\bibfnamefont {J.~G.}\ \bibnamefont
  {White}}, \bibinfo {author} {\bibfnamefont {E.}~\bibnamefont {Southgate}},
  \bibinfo {author} {\bibfnamefont {J.~N.}\ \bibnamefont {Thomson}},\ and\
  \bibinfo {author} {\bibfnamefont {S.}~\bibnamefont {Brenner}},\ }\bibfield
  {title} {\bibinfo {title} {The structure of the nervous system of the
  nematode {{Caenorhabditis}} elegans},\ }\href
  {https://doi.org/10.1098/rstb.1986.0056} {\bibfield  {journal} {\bibinfo
  {journal} {Philos. Trans. Royal Soc. B}\ }\textbf {\bibinfo {volume} {314}},\
  \bibinfo {pages} {1} (\bibinfo {year} {1986})}\BibitemShut {NoStop}%
\bibitem [{\citenamefont {Karrer}\ \emph {et~al.}(2014)\citenamefont {Karrer},
  \citenamefont {Newman},\ and\ \citenamefont {Zdeborov\'a}}]{internet_as}%
  \BibitemOpen
  \bibfield  {author} {\bibinfo {author} {\bibfnamefont {B.}~\bibnamefont
  {Karrer}}, \bibinfo {author} {\bibfnamefont {M.~E.~J.}\ \bibnamefont
  {Newman}},\ and\ \bibinfo {author} {\bibfnamefont {L.}~\bibnamefont
  {Zdeborov\'a}},\ }\bibfield  {title} {\bibinfo {title} {Percolation on sparse
  networks},\ }\href {https://doi.org/10.1103/PhysRevLett.113.208702}
  {\bibfield  {journal} {\bibinfo  {journal} {Phys. Rev. Lett.}\ }\textbf
  {\bibinfo {volume} {113}},\ \bibinfo {pages} {208702} (\bibinfo {year}
  {2014})}\BibitemShut {NoStop}%
\bibitem [{\citenamefont {{University of Oregon Route Views
  Project}}(2001)}]{route_views_bgp_data}%
  \BibitemOpen
  \bibfield  {author} {\bibinfo {author} {\bibnamefont {{University of Oregon
  Route Views Project}}},\ }\href@noop {} {\bibinfo {title} {Bgp as internet
  traffic data - oregon route views project}},\ \bibinfo {howpublished}
  {\url{http://www.routeviews.org/}} (\bibinfo {year} {2001}),\ \bibinfo {note}
  {data spanning from 8 November 1997 to 2 January 2000. Collected by
  NLANR/MOAT.}\BibitemShut {Stop}%
\end{thebibliography}
%

\end{document}